\documentclass[preprint]{aastex61}

\usepackage{amsmath}
\usepackage{graphicx}
\usepackage{natbib}
\usepackage{ulem}
\usepackage{bm}

\usepackage{xspace}
\usepackage{hyperref}

\usepackage{multirow}

\newcommand{\rcminus}{\ensuremath{r_{\rm c}^{-}}}
\newcommand{\rcplus}{\ensuremath{r_{\rm c}^{+}}}

\newcommand{\rAminus}{\ensuremath{r_{\rm A}^{-}}}
\newcommand{\rAplus}{\ensuremath{r_{\rm A}^{+}}}

\newcommand{\hatFA}{\ensuremath{\hat{F}_{\rm A}}}
\newcommand{\hatFc}{\ensuremath{\hat{F}_{\rm c}}}

\newcommand{\Ampere}{Amp$\acute{\rm e}$re}

\newcommand{\sideal}{\ensuremath{s_{\rm ideal}}}
\newcommand{\sres}{\ensuremath{s_{\rm res}}}
\newcommand{\Sideal}{\ensuremath{S_{\rm ideal}}}
\newcommand{\Sres}{\ensuremath{S_{\rm res}}}

\newcommand{\lRmin}{\ensuremath{(l/R)_{\rm min}}}
\newcommand{\tauPmin}{\ensuremath{(\tau/P)_{\rm min}}}

\newcommand{\ptot}{\ensuremath{p_{\rm tot}}}
\newcommand{\ptottilde}{\ensuremath{\tilde{p}_{\rm tot}}}
\newcommand{\Alf}{Alfv$\acute{\rm e}$n}
\newcommand{\Rm}{\ensuremath{R_{\rm m}}}
\newcommand{\rc}{\ensuremath{r_{\rm c}}}
\newcommand{\rA}{\ensuremath{r_{\rm A}}}

\newcounter{RomanNumber}

\def\myvect#1{\ensuremath{\bm{#1}}}

\newcommand{\kms}{\ensuremath{{\rm km}~{\rm s}^{-1}}}

\newcommand{\mui}{\ensuremath{\mu_{\rm i}}}
\newcommand{\mue}{\ensuremath{\mu_{\rm e}}}

\newcommand{\cs}{\ensuremath{c_{\rm s}}}
\newcommand{\csi}{\ensuremath{c_{\rm si}}}
\newcommand{\cse}{\ensuremath{c_{\rm se}}}

\newcommand{\va}{\ensuremath{v_{\rm A}}}
\newcommand{\vai}{\ensuremath{v_{\rm Ai}}}
\newcommand{\vae}{\ensuremath{v_{\rm Ae}}}

\newcommand{\ct}{\ensuremath{c_{\rm T}}}
\newcommand{\cti}{\ensuremath{c_{\rm Ti}}}
\newcommand{\cte}{\ensuremath{c_{\rm Te}}}

\newcommand{\omgR}{\ensuremath{\omega_{\rm R}}}
\newcommand{\omgI}{\ensuremath{\omega_{\rm I}}}

\newcommand{\vph}{\ensuremath{v_{\rm ph}}}

\newcommand{\rhoi}{\ensuremath{\rho_{\rm i}}}
\newcommand{\rhoe}{\ensuremath{\rho_{\rm e}}}

\newcommand{\mathd}{\ensuremath{{\rm d}}}

\submitjournal{ApJ}

\shorttitle{Waves in photospheric waveguides}
\shortauthors{Chen et al.}
\begin{document}

\title{Damping of slow surface kink modes 
   in solar photospheric waveguides modeled by
   one-dimensional inhomogeneities}

\correspondingauthor{Bo Li}
\email{bbl@sdu.edu.cn}

\author{Shao-Xia Chen}
\affiliation{Shandong Provincial Key Laboratory of Optical Astronomy and Solar-Terrestrial Environment,
   Institute of Space Sciences, Shandong University, Weihai 264209, China}

\author{Bo Li}
\affiliation{Shandong Provincial Key Laboratory of Optical Astronomy and Solar-Terrestrial Environment,
   Institute of Space Sciences, Shandong University, Weihai 264209, China}

\author{Tom Van Doorsselaere}
\affiliation{Centre for mathematical Plasma Astrophysics (CmPA), KU Leuven, Celestijnenlaan 200B bus 2400, B-3001 Leuven, Belgium}

\author{Marcel Goossens}
\affiliation{Centre for mathematical Plasma Astrophysics (CmPA), KU Leuven, Celestijnenlaan 200B bus 2400, B-3001 Leuven, Belgium}

\author{Hui Yu}
\affiliation{Shandong Provincial Key Laboratory of Optical Astronomy and Solar-Terrestrial Environment,
   Institute of Space Sciences, Shandong University, Weihai 264209, China}

\author{Micha{\"e}l Geeraerts} 
\affiliation{Centre for mathematical Plasma Astrophysics (CmPA), KU Leuven, Celestijnenlaan 200B bus 2400, B-3001 Leuven, Belgium}



\begin{abstract}
Given the recent interest in magnetohydrodynamic (MHD) waves in pores and sunspot umbrae, we examine the damping of slow surface kink modes (SSKMs) by modeling solar photospheric waveguides with a cylindrical inhomogeneity comprising a uniform interior, a uniform exterior, and a continuous transition layer (TL) in between. Performing an eigen-mode analysis in linear, resistive, gravity-free MHD, our approach is idealized in that, among other things, our equilibrium is structured only in the radial direction. We can nonetheless address two damping mechanisms simultaneously, one being the Ohmic resistivity, and the other being the resonant absorption of SSKMs in the cusp and Alfv$\acute{\rm e}$n continua. We find that the relative importance of the two mechanisms  depends sensitively on the magnetic Reynolds number ($R_{\rm m}$). Resonant absorption is the sole damping mechanism for realistically large values of $R_{\rm m}$, and the cusp resonance in general dominates the Alfv$\acute{\rm e}$n one unless the axial wavenumbers are at the lower end of the observationally relevant range. We also find that the thin-boundary approximation holds only when the TL-width-to-radius ratios are much smaller than nominally expected. The Ohmic resistivity is far more important for realistically small $R_{\rm m}$. Even in this case, SSKMs are only marginally damped, with  damping-time-to-period-ratios reaching $\sim 10$ in the parameter range we examine. 
\end{abstract}



\section{INTRODUCTION}
\label{sec_intro}

Low-frequency waves and oscillations in the magnetohydrodynamic (MHD) regime     
    often prove key in
    the gas-magnetic field interactions throughout
    the solar atmosphere \citep[see e.g., the textbook by][]{roberts_2019CUP}.
When generated by, say, sub-photospheric convective motions,
    these waves may be sufficiently energetic to play an active role in atmospheric heating,
    thereby helping shape such atmospheric structures as coronal loops
    \citep[see e.g.,][for recent reviews]{2012RSPTA.370.3217P,2015RSPTA.37340261A,2019ARA&A..57..157C}.
Even when not that energetic, these waves can passively
    encode some rich information
    on their host magnetic structures in their observational signatures.
Conversely, this information can be deciphered 
    with increasingly refined theories of MHD waves in an inhomogeneous medium,
    constituting the field of ``coronal seismology'' or more generally
    ``solar atmospheric seismology''
    (SAS, 
    \citeauthor{1984ApJ...279..857R}~\citeyear{1984ApJ...279..857R};
    see also recent reviews by e.g., 
    \citeauthor{2000SoPh..193..139R}~\citeyear{2000SoPh..193..139R},
    \citeauthor{2005LRSP....2....3N}~\citeyear{2005LRSP....2....3N},
    \citeauthor{2020ARAA..Nakariakov}~\citeyear{2020ARAA..Nakariakov}).
Whichever the application, MHD waves have been customarily classified
    with a scheme largely established for a canonical configuration where
    a magnetic structure is seen as
    a field-aligned plasma cylinder with circular cross-section,
    and the equilibrium quantities are structured only in the radial direction
	(\citeauthor{1979A&A....76...20W}~\citeyear{1979A&A....76...20W},  
	\citeauthor{1982SoPh...75....3S}~\citeyear{1982SoPh...75....3S},
	\citeauthor{1983SoPh...88..179E}~\citeyear{1983SoPh...88..179E}, hereafter ER83;
	see also \citeauthor{1970A&A.....9..159R}~\citeyear{1970A&A.....9..159R},
	\citeauthor{1975IGAFS..37....3Z}~\citeyear{1975IGAFS..37....3Z},
	\citeauthor{1986SoPh..103..277C}~\citeyear{1986SoPh..103..277C}).
We restrict ourselves to collective waves, 
    ``collective'' in the sense that they involve 
    coherent motions of the fluid parcels in the structure itself
    and possibly its surroundings. 
Collective waves (``modes" hereafter) may be classified as 
    sausage ($m=0$), kink ($m=1$), and fluting modes ($m\ge 2$)
    by their azimuthal wavenumber $m$.
They may be independently grouped
   into slow and fast modes according to their axial phase speeds.
Alternatively, the spatial dependence of the associated perturbations in the surroundings
    enables the notions of trapped and leaky modes 
    as a measure for classification. 
Likewise, the notions of surface and body modes 
    are enabled by an examination of the spatial behavior
    within the magnetic structure.
One therefore ends up with an extensive list of combinations like
    fast kink body modes and slow sausage surface modes
\footnote{
    Some issue arises when one applies this classification scheme to the
       abundantly observed kink modes that are trapped in thin coronal loops
       and that possess axial phase speeds comparable to the \Alf\ speed.
    Customarily called ``fast kink body modes'', they turn out to be essentially
       {\Alf}ic surface waves with mixed properties
       \citep[for details, see][]{2009A&A...503..213G,2012ApJ...753..111G}.
    In particular, they survive in incompressible MHD, 
       making them difficult to be classified as ``fast''.
	We therefore deem it more appropriate to call
	   them ``the radial fundamental kink mode''.
    }.
In the structured solar corona, 
    there have been ample observations for modes
    pertaining to a substantial fraction of 
    these combinations
    \citep[see the reviews by e.g.,][]{2007SoPh..246....3B,2016SSRv..200...75N,2016GMS...216..395W}.
In particular, the ubiquity of radial fundamental kink modes has enabled one
    to seismologically
    construct the spatial distributions of the magnetic field strength 
    in a substantial portion of some active region
    (AR, \citeauthor{2019ApJ...884L..40A}~\citeyear{2019ApJ...884L..40A})
    or even across several ARs \citep{2020arXiv200803136Y}.

There is a long history of
   observational studies on MHD waves in the lower solar atmosphere as well
   (see e.g., the review by
     \citeauthor{2015LRSP...12....6K}~\citeyear{2015LRSP...12....6K} on waves in sunspots,
    and that by 
     \citeauthor{2015SSRv..190..103J}~\citeyear{2015SSRv..190..103J} on waves in the chromosphere).
However, it seems that the classification of these waves with the ER83 scheme 
    was made possible only relatively recently. 
Take the quiet chromosphere for instance.   
Cyclic transverse displacements, the defining characteristic of radial fundamental kink modes,
    have been abundantly observed in slender fibrils imaged in both
    Ca II 3969~\AA\ with SUNRISE/SuFI \citep{2017ApJS..229....9J}
    and H$_{\alpha}$ with ROSA \citep{2012NatCo...3.1315M}. 
Likewise, the latter study also indicated the omnipresence of breathing-like motions
    of the H$_{\alpha}$ fibrils, thereby suggesting the ubiquity
    of fast sausage modes.      
In both studies, the width of chromospheric fibrils reaches down to 
    $\sim 150$~km, which highlights the importance of 
    high spatial resolution in enabling such classifications as 
    kink or sausage modes. 
The same can be said for waves in chromospheres above sunspot umbrae.
For instance, the long-known spiral wave patterns (SWPs) with periodicities 
    of $\sim 3$~mins \citep[e.g.,][and references therein]{2016ApJ...817..117S}
    were recently shown to be compatible with the superposition of 
    slow body modes with $m=0$ and $m=1$ ($m=0$ and $m=2$)
    in the case of one-armed (two-armed) SWPs~\citep{2019ApJ...877L...9K}. 
Likewise, clear signatures of slow kink body modes were identified
    in the temporally and spatially filtered H$_{\alpha}$ image sequences
    acquired with ROSA mounted on the Dunn Solar Telescope \citep[DST,][]{2017ApJ...842...59J}.
Moving to the photosphere in either sunspot umbrae or pores, 
    we note that slow sausage modes tend to be ubiquitous
    \citep[e.g.,][]{2008IAUS..247..351D, 2009ApJ...702.1443F,2011ApJ...729L..18M,
        2014A&A...563A..12D,2015A&A...579A..73M,2015ApJ...806..132G,2016ApJ...817...44F}.
In addition, surface modes tend to be favored over body modes in pores with larger
    sizes and/or stronger magnetic fields  
    (\citeauthor{2018ApJ...857...28K}~\citeyear{2018ApJ...857...28K},
    \citeauthor{2020arXiv200711594G}~\citeyear{2020arXiv200711594G};
    see also \citeauthor{2018ApJ...869..110S}~\citeyear{2018ApJ...869..110S}).
The waves in the lower solar atmosphere, be them kink or sausage modes,
    were suggested to reach chromospheric levels with a Poynting flux of up to 
    $\sim 27~{\rm kW}~{\rm m}^{-2}$~\citep[e.g.,][]{2009ApJ...702.1443F},
    a value sufficiently large to meet the chromospheric heating requirement.
Equally useful are their applications to SAS as initiated by
    \citet{2009ApJ...702.1443F} and \citet{2015A&A...579A..73M}.
Observations aside, theoretical studies have proven crucial for establishing
    the phase-differences between relevant perturbations and therefore for
    aiding mode identification~\citep{2013A&A...551A.137M,2013A&A...555A..75M}.
Likewise, expressions for the energy and energy flux densities are inevitable 
    for examining the energetics~\citep{2015A&A...578A..60M}.          
         
This study is intended to examine the damping of slow surface kink modes (SSKMs) 
    in photospheric waveguides representative of pores and sunspot umbrae, thereby
    partially addressing the fate of SSKMs generated by, 
    say, sub-photospheric convective motions. 
Our motivation is twofold.
Firstly, in contrast to the apparent omnipresence of slow surface sausage modes (SSSMs),
     photospheric kink modes were implicated only
     in the Hinode/SP measurements reported by \citet{2009ApJ...702.1443F} 
     to our knowledge, and the kink mode therein is more likely to
     be a fast surface one. 
We take this as encouraging rather than discouraging for a theoretical study on
     SSKMs, the reason primarily related to their theoretical properties 
     and their likely excitation. 
Theoretically, SSKMs tend to be indistinguishable from 
     SSSMs in terms of eigen-frequencies
     (see e.g., Figure~3 in ER83).      
While the azimuthal variations of the two modes are distinctively different, 
     the random broadband motions that excite SSSMs are likely to excite
     SSKMs as well. 
Secondly, SSSMs in pores were observationally demonstrated to be heavily damped over 
     photospheric heights~\citep{2015ApJ...806..132G,2020arXiv200711594G}.
So far this heavy damping has only been partially addressed.
Working in the thin-boundary (TB) approximation, the largely analytical study
     by \citet{2017A&A...602A.108Y} indicated that the resonant absorption
     of SSSMs in the cusp continuum may result in a damping-time-to-period ratio
     ($\tau/P$) of $\sim 10$.       
This was taken further by \citet{2019ApJ...879..121S} who incorporated magnetic twist
     in the equilibrium.
As such SSSMs further resonantly couple to the Alfv\'en continuum,
     thereby leading to some stronger damping. 
Note that the \Alf\ and cusp continua arise when the piece-wise constant model
     is replaced by a model
     in which the equilibrium quantities vary in a continuous manner in a transitional
     layer (TL) from their internal to their external values. 
By construction, the TB results apply only when
     the TL width ($l$) is far below
     the cylinder radius ($R$). 
However, in the above-referenced TB studies,
     strong damping tends to occur when $l/R$ is substantial, and one may question whether the TB damping rates hold.
Indeed, focusing on the damping of SSSMs due to the cusp resonance,
     \citet[][hereafter C18]{2018ApJ...868....5C} showed 
     with resistive MHD computations 
     that the numerically derived damping rates are in general 
     well below the TB values found by \citet{2017A&A...602A.108Y}.  
C18 further showed that in general Ohmic resistivity tends to be more important
     than resonant absorption for damping SSSMs, a result that was further corroborated 
     by the analytical study in \citet[][G20 from here on]{2020ApJ...897..120G}. 
Compared with SSSMs, SSKMs are theoretically appealing in that
     they can resonantly couple to the $m=1$ Alfv\'en waves even in the absence of 
     magnetic twist \citep[][Y17 hereafter]{2017ApJ...850...44Y}.
In Y17, the damping of SSKMs was solely due to resonant absorption and
     the cusp resonance
     was shown to dominate for large axial wavenumbers, resulting in values
     of down to $\sim 10$ for $\tau/P$.  
Now two questions arise,      
     given that Y17 adopted the TB approximation, and given
     the C18 results regarding SSSMs. 
Firstly, how well do the Y17 results apply as far as
     the damping due to resonant absorption is concerned?
Secondly, what is the role of Ohmic resistivity in the
     damping of SSKMs?

Some remarks prove necessary at this point. 
We will adopt resistive, gravity-free, MHD as our theoretical framework,  
   in accordance with which we will additionally take our equilibrium configuration
   to be structured only transversely. 
Both call into question how well our results apply to photospheric structures
   in reality, the structuring of which is way more complicated than assumed here.
We leave the discussions on this aspect till the end of this manuscript.
It suffices to note here that by ``photospheric modes" we refer to
   the modes supported by some idealized magnetic structures, for which the ordering of
   the internal and external characteristic speeds nonetheless captures some
   realistic features of pores and sunspot umbrae.        
Bearing the caveats in mind, we list two features that make our study 
   considerably new relative to available ones on photospheric modes. 
Firstly, so far the damping of SSKMs was examined only by Y17 to our knowledge.
Relative to Y17, our study is new
   in that we will simultaneously account for the effects of
   the Ohmic resistivity and resonant absorption, whereas
   Y17 considered only resonant absorption.  
Secondly, we will analyze the wave damping from the
   perspective of energy balance, whereas all the afore-referenced studies approached the wave damping exclusively 
   from the perspective of eigen-frequencies. 
As we will acknowledge where more appropriate,    
   this is not to say that wave damping has not been examined
   from the energetics perspective.
It is just that ``photospheric modes'' have not been examined this way as far as
   we know.

This manuscript is organized as follows. 
Section~\ref{sec_model} 
    details the specification of our equilibrium configuration,
    and how we formulate and numerically solve the pertinent eigenvalue problem.
The numerical results are then described in Section~\ref{sec_num_results}. 
Section~\ref{sec_conc} summarizes the present study, ending 
    with some concluding remarks.              

\section{Model Description}
\label{sec_model}

\subsection{Equilibrium Configuration}
\label{sec_equilibrium}
We adopt resistive, gravity-free MHD throughout, 
    for which the primitive variables are the mass density
    ($\rho$), velocity ($\myvect{v}$), magnetic field ($\myvect{B}$), 
    and thermal pressure ($p$). 
We denote the equilibrium quantities with the subscript $0$, and assume that
    no equilibrium flow is present ($\myvect{v}_0 = 0$). 
Working in a cylindrical coordinate system $(r, \theta, z)$, we assume that
    all equilibrium quantities depend only on $r$. 
The equilibrium magnetic field ($\myvect{B}_0$) is taken to be in the $z$-direction. 
As such $B_0(r)$ and $p_0(r)$ are related by the transverse force balance condition
\begin{eqnarray}
\label{eq_freedom}
\displaystyle
  p_0(r) + \frac{B_0(r)^2}{2\mu_0}={\rm const}~,
\end{eqnarray}
    where $\mu_0$ is the magnetic permeability in free space. 
The adiabatic sound speed $\cs$, 
    \Alf\ speed $\va$, and cusp (tube) speed $\ct$ are then defined by 
\begin{eqnarray}
\label{eq_speeds}
\displaystyle
  \cs^2(r) = \frac{\gamma p_0}{\rho_0}~,~~~
  \va^2(r) = \frac{B_0^2}{\mu_0\rho_0}~,~~~
  \ct^2(r) = \frac{\cs^2 \va^2}{\cs^2+\va^2}~,
\end{eqnarray}
    with $\gamma=5/3$ being the adiabatic index.
    
We model a photospheric waveguide as a field-aligned cylinder
    with mean radius $R$.     
Following C18, we realize this by assuming that $\cs^2$ and $\ct^2$
    take the following form
 \begin{eqnarray}
\displaystyle
{\cal E}(r)=\left\{
\begin{array}{lc}
{\cal E}_{\rm i}~,& 0<r<r_{\rm i}=R-l/2~; \\[0.2cm]
    \displaystyle\frac{{\cal E}_{\rm i}+{\cal E}_{\rm e}}{2}
   -\frac{{\cal E}_{\rm i}-{\cal E}_{\rm e}}{2}
     \sin\frac{\pi(r-R)}{l}~,& r_{\rm i}\le r\le r_{\rm e}=R+l/2~;	\\[0.2cm]	
     {\cal E}_{\rm e}~, & r>r_{\rm e}~,		
\end{array}
\right.
\label{eq_profile_cs_ct}
\end{eqnarray}
    where ${\cal E}$ represents $\cs^2$ and $\ct^2$.
In addition, the subscripts~i and e refer to 
    the equilibrium values at the cylinder axis and in the ambient medium, respectively.
As such the equilibrium configuration comprises a uniform interior ($r < r_{\rm i}$),
    a uniform exterior ($r > r_{\rm e}$) and a transition layer (TL) continuously
    connecting the two.
This equilibrium is fully specified by a set of dimensional parameters
    $[\rhoi, \vai, R]$ together with a set of dimensionless parameters
    $[\csi/\vai, \cse/\vai, \vae/\vai, l/R]$.      
As in C18, we fix     
    $[\csi, \cse, \vae]$ at $[0.5, 0.75,  0.25] \vai$, 
    which is identical to ER83 and also in agreement with
    the generally accepted values for sunspot umbrae and pores \citep[e.g.,][]{2017ApJ...842...59J,2017ApJ...850...44Y}.
The internal and external cusp speeds then read
    $[\cti, \cte] = [0.4472, 0.2372]~\vai$. 
The ratio of the TL width to the mean radius ($l/R$),
    on the other hand, is allowed to vary.     
The spatial distributions for both the characteristic speeds
    and primitive variables are displayed in Figure~\ref{fig_EQprofile}, 
    where $l/R$ is taken to be $0.2$ for illustration purposes.     
Among the dimensional parameters, $\vai$ and $R$ are relevant for our purposes.
We take $\vai = 12~\kms$ in view of Figure~4 in \citet{2015ApJ...806..132G},
    who in turn constructed the values with the ``hot'' umbral model
    from \citet{1986ApJ...306..284M}.
With pores and sunspot umbrae in mind, we assume that $R$ ranges from 
    $500$~km 
    (the lower limit for pores, see \citeauthor{2003AN....324..369S}~\citeyear{2003AN....324..369S})           
    to $10^4$~km
    (for large sunspot umbrae, see Figure~1 in 
    \citeauthor{2018ApJ...869..110S}~\citeyear{2018ApJ...869..110S}).
    
\subsection{Formulation of the Eigenvalue Problem and Method of Solution}
\label{sec_linearMHD}

Let the subscript $1$ denote small-amplitude perturbations to the equilibrium.     
The linearized resistive MHD equations then read 
\begin{eqnarray}
 \label{eq_MHD}
  \displaystyle
   \rho_0 \frac{\partial \myvect{v}_{1}}{\partial t}
     &=& -\nabla p_1 +\frac{(\nabla\times \myvect{B}_0)\times \myvect{B}_1}{\mu_0}
                   +\frac{(\nabla\times\myvect{B}_1)\times \myvect{B}_0}{\mu_0}~,
                   \label{eq_linMHD_momen} \\ [0.2cm]
  \displaystyle
   \frac{\partial \myvect{B}_1}{\partial t}
     &=& \nabla\times\left(\myvect{v}_1\times \myvect{B}_0-\frac{\eta}{\mu_0}\nabla\times\myvect{B}_1
                   \right)~, \label{eq_linMHD_Farad} \\[0.2cm]
  \displaystyle
   \frac{\partial p_1}{\partial t}
     &=& -\myvect{v}_1\cdot\nabla p_0 -\gamma p_0 \nabla\cdot \myvect{v}_1
       + \frac{2(\gamma-1)\eta}{\mu_0^2}
            \left(\nabla\times \myvect{B}_1\right)\cdot\left(\nabla\times \myvect{B}_0\right)~.
            \label{eq_linMHD_pres}
\end{eqnarray}
Here $\eta$ is the Ohmic resistivity, taken to be constant for simplicity.
We adopt an eigen-value-problem (EVP) perspective by  
    Fourier-analyzing any perturbation as
\begin{eqnarray}
\label{eq_Fourier}
   g_1 (r, \theta, z; t)
 = {\rm Re}\{\tilde{g}(r)\exp[-i(\omega t-m\theta-kz)]\}~,
\end{eqnarray}
   where $\omega$ is the angular frequency, and $k$ ($m$)
   represents the axial (azimuthal) wavenumber.
With tilde we denote the Fourier amplitude.         
We take $k$ as real-valued, but allow $\omega$ to be complex-valued. 
If some quantity is complex, then we denote
    its real and imaginary parts with subscripts~R and I, respectively.
It follows that
    the period $P = 2\pi/\omgR$, and 
    the damping time $\tau = 1/|\omgI|$.
Instabilities are not of interest, hence $\omgI < 0$.    
The axial phase speed is defined by 
    $\vph = \omgR/k$.
   
We proceed with normalizing Equations~\eqref{eq_linMHD_momen}
   to \eqref{eq_linMHD_pres}. 
Among the necessary normalizing constants,     
   we take $R$, $\vai$, and $\rhoi$ as independent ones,
   following from which are the normalizing constants for time ($R/\vai$),
   the
   magnetic field ($B_{\rm i} \equiv \sqrt{\mu_0 \rhoi \vai^2}$)
   and thermal pressure ($\rhoi\vai^2 = B_{\rm i}^2/\mu_0$). 
The Ohmic resistivity ends up in the magnetic Reynolds number 
   $\Rm = \mu_0 R \vai/\eta$.     
In component form, the resulting dimensionless equations then read
 \begin{eqnarray}
%\begin{tiny}
\omega\tilde{v}_r & = & \displaystyle
     -\frac{B_0}{\rho_0}\left(k\tilde{B}_r +i\frac{{\rm d}\tilde{B}_z}{{\rm d}r}\right)
     -\frac{i \tilde{B}_z}{\rho_0}\frac{{\rm d}B_0}{{\rm d}r}
     -\frac{i}{\rho_0}\frac{{\rm d}\tilde{p}}{{\rm d}r}~,  \label{eq_Fourier_vr}\\[0.2cm]
\omega\tilde{v}_\theta & = & \displaystyle
     -\frac{B_0}{\rho_0}\left(k\tilde{B}_\theta -\frac{m\tilde{B}_z}{r}\right)
     +{}\frac{m\tilde{p}}{r\rho_0}~,  \label{eq_Fourier_vtheta}\\[0.2cm]
\omega\tilde{v}_z & = & \displaystyle
      \frac{i \tilde{B}_r}{\rho_0}\frac{{\rm d}B_0}{{\rm d}r}
     +\frac{k\tilde{p}}{\rho_0}~, 			   \label{eq_Fourier_vz}	\\[0.2cm]
\omega\tilde{B}_r & = & \displaystyle
     -kB_0\tilde{v}_r +
     \frac{1}{R_{\rm m}}\left(i\frac{{\rm d^2}\tilde{B}_r}{{\rm d}{r}^2}
         +\frac{i}{r}\frac{{\rm d}\tilde{B}_r}{{\rm d}r}
         -\frac{im^2\tilde{B}_r}{{r}^2}
         -i{k}^2\tilde{B}_r+\frac{2m\tilde{B}_\theta}{r^2}
         -\frac{i\tilde{B}_r}{{r}^2}
     \right)~, 						   \label{eq_Fourier_Br} \\[0.2cm]
\omega\tilde{B}_\theta & = & \displaystyle
     -kB_0\tilde{v}_\theta +
     \frac{1}{R_{\rm m}}\left(i\frac{{\rm d^2}\tilde{B}_\theta}{{\rm d}{r}^2}
         +\frac{i}{r}\frac{{\rm d}\tilde{B}_\theta}{{\rm d}r}
         -\frac{im^2\tilde{B}_\theta}{{r}^2}
         -i{k}^2\tilde{B}_\theta-\frac{2m\tilde{B}_r}{r^2}
         -\frac{i\tilde{B}_\theta}{{r}^2}
     \right)~, 						   \label{eq_Fourier_Btheta} \\[0.2cm]
\omega\tilde{B}_z & = & \displaystyle
    -i\tilde{v}_r\frac{{\rm d}B_0}{{\rm d}r}
    -B_0\left(i\frac{{\rm d}\tilde{v}_r}{{\rm d}r}+i\frac{\tilde{v}_r}{r}
    -\frac{m\tilde{v}_\theta}{r}\right)
    +\frac{1}{R_{\rm m}}\left(i\frac{{\rm d^2}\tilde{B}_z}{{\rm d}{r}^2}
         +\frac{i}{r}\frac{{\rm d}\tilde{B}_z}{{\rm d}r}
         -\frac{im^2\tilde{B}_z}{{r}^2}-i{k}^2\tilde{B}_z
     \right)~,						   \label{eq_Fourier_Bz}	\\[0.2cm]
\omega\tilde{p} & = & \displaystyle
    -c_{\rm s}^2\rho_0
    \left(i\frac{{\rm d}\tilde{v}_r}{{\rm d}r}
         +i\frac{\tilde{v}_r}{r}-\frac{m\tilde{v}_\theta}{r}-k\tilde{v}_z
       \right)
    -i\tilde{v}_r\frac{{\rm d}p_0}{{\rm d}r}
    +\frac{2\left(\gamma-1\right)}{R_{\rm m}}\frac{{\rm d}B_0}{{\rm d}r}
       \left(k\tilde{B}_r+i\frac{{\rm d}\tilde{B}_z}{{\rm d}r}\right)~.
         \label{eq_Fourier_p}
%\end{tiny}
\end{eqnarray}
We specialize to the case where $m=1$, given that SSKMs are of interest. 
Equations~\eqref{eq_Fourier_vr} to \eqref{eq_Fourier_p}
    constitute a standard EVP when supplemented with
    appropriate boundary conditions (BCs).
The BCs at the cylinder axis ($r=0$) are that
    $\mathd\tilde{v}_r/\mathd r = \mathd\tilde{v}_\theta/\mathd r
    =\mathd\tilde{B}_r/\mathd r = \mathd\tilde{B}_\theta/\mathd r
    =\tilde{v}_z = \tilde{B}_z = \tilde{p} 
    =0$. 
With trapped modes in mind, we require that all variables 
    vanish when $r$ approaches infinity.  
From here onward, we see the physical variables as dimensional again
    for the ease of description.       

The EVP is formulated and solved 
    with the general-purpose finite-element code PDE2D~\citep{1988Sewell_PDE2D},
    which was first introduced to the solar context
    by \citet{2006ApJ...642..533T} to our knowledge.
The computational domain spans from $0$ to some outer boundary $r_{\rm M}$.
A nonuniform grid is adopted, among which
    a considerable number of grid points are deployed
    in the TL to resolve the possible oscillatory behavior therein.
The total number of grid points typically ranges from 
    $5\times 10^4$ up to $2.5\times 10^5$,
    among which up to $50\%$ are deployed
    in the TL in some cases.
Overall the strategy is that deploying more grid points in the TL leads
    to no discernible difference to our results, by which we mean
    both the eigen-frequencies and eigen-functions.    
The outer boundary is placed at a sufficiently large distance of $r_{\rm M} = 50~R$
    to ensure that further increasing $r_{\rm M}$ does not introduce any appreciable change to our results. 
Our numerical results are remarkably accurate, as will be seen from
    the energetics examinations.
Taking $[l/R, kR, \Rm]$ as free parameters, one may formally 
    express $\omega$ as
\begin{eqnarray}
    \frac{\omega R}{v_{\rm Ai}} = {\cal G}\left(kR, \frac{l}{R}, R_{\rm m}\right)~.
\label{eq_omega_formal}
\end{eqnarray}
With $l/R$ difficult to constrain observationally,
    we allow $l/R$ to vary as broadly as the code permits. 
Evaluating $\Rm$ necessarily involves the electric resistivity, for which purpose
    we quote the values for 
        electric conductivity listed in Table~2 of 
        \citet{1983SoPh...84...45K} for a model sunspot umbra in the temperature range of $3500$ to $\sim 6400$~K.    
Given the range of $R$ we have discussed and the internal \Alf\ speed we have fixed,
    $\Rm$ then ranges from $10^4$ to $6.6\times 10^7$.

The range of $kR$ is difficult to constrain,
    given that
    a concrete identification of SSKMs has yet to appear.
We nonetheless assume that $kR \gtrsim 0.3$ with the limited knowledge
    on the axial wavelength ($\lambda = 2\pi/k$) of
    photospheric modes.
This adopted range of $kR$ is inferred primarily from the measurements
	of photospheric slow sausage modes.
Gathering the measurements for pores and sunspot umbrae of various sizes, 
	one finds that the period tends to range between $\sim 1.5$ and $20$~mins
	\citep{2009ApJ...702.1443F,2011ApJ...729L..18M,2015A&A...579A..73M,2015ApJ...806..132G,2016ApJ...817...44F, 2018ApJ...857...28K, 2020arXiv200711594G}. 
Boldly assuming that this periodicity applies to SSKMs in our equilibrium, 
	we find that $\lambda$ ranges from $\sim 500$ to $\sim 6500$~km by approximating
	the axial phase speed $\vph$ with $\cti$.
We gather further information from      
	the spectropolarimetric measurements with 
	Fe I 6173~\AA\ and Ca II 8542~\AA\ for a large sunspot 
	as reported by \citet{2018ApJ...869..110S}.
The phase delay between the oscillatory signals in these two lines 
	was found to peak at $\delta\phi \approx 0.5$~rad
	(Figure~3 therein).
	As suggested by the same study, 
	the difference between the formation heights of the two lines ($d$)
	is $\sim 800$~km.
Equating $2\pi d/\lambda$ to $\delta\phi$ leads to a $\lambda \approx 10^4$~km.
We therefore deem $\lambda$ to range from $500$ to $10^4$~km, from which
	we then deduce that $kR \gtrsim 0.3$.  
This deduction is admittedly idealized.
To name but one caveat, in general one should 
	equal $2\pi d/\lambda$ to $2n\pi + \delta\phi$ ($n=0, 1, 2, \cdots$).
Letting $n=1$, one finds a $\lambda$ of $\sim 740$~km, which is
	drastically different from the case where $n=0$. 
This subtlety is representative of multi-wavelength studies on waves 
	in the lower solar atmosphere.

It is necessary to address some technical details regarding
    Equation~\eqref{eq_omega_formal}, for which purpose
    we recall that only the parameters in the parentheses are seen as adjustable.
The ER83 results were established in ideal MHD ($\eta \equiv 0$) 
    and for a step transverse distribution ($l=0$).
In this case, an SSKM can be identified without ambiguity for a given $kR$.
However, care needs to be exercised in resistive MHD ($\eta \ne 0$), 
    in which case a dense set of solutions arise
    (see e.g., Figure~2 in 
    \citeauthor{2020ApJ...897..120G}~\citeyear{2020ApJ...897..120G}
    even though SSSMs were examined therein).
We would like to find a unique solution out of this dense set, 
    ``unique'' in the sense that it falls back to a standard
    SSKM as in ER83.
As in C18, for a target combination $[kR, l/R, \Rm]$,
    this is done by first fixing $[kR, l/R]$ and then
         looking for the solution lying on the curve 
         that becomes independent on $\Rm$ when $\Rm$ is sufficiently large.
This procedure relies on two facts,
    one being that the $\Rm$-independent eigen-frequencies 
        pertain to resonantly damped collective modes,
    and the other being that resonantly damped modes
        are physically connected to their ER83 counterparts
    (see the review by \citeauthor{2011SSRv..158..289G}~\citeyear{2011SSRv..158..289G};
    see also \citeauthor{1991PhRvL..66.2871P}~\citeyear{1991PhRvL..66.2871P},
    and \citeauthor{2004ApJ...606.1223V}~\citeyear{2004ApJ...606.1223V}).

\subsection{Resonantly Damped SSKMs in the Thin-Boundary Limit}
\label{sec_TB}
When $\Rm$ is irrelevant in Equation~\eqref{eq_omega_formal}, 
   the functional dependence therein can be established 
   with a largely analytical approach
   in the TB limit ($l/R \ll 1$). 
In this case, the eigen-frequencies for resonantly damped SSKMs are 
   governed by the dispersion relation (DR) 
   derived by Y17, which reads
\begin{eqnarray}
\label{eq_TB}
\begin{array}{rcl}
&& 
\displaystyle 
    \rhoi (\omega^2-k^2 \vai^2) 
  - \rhoe (\omega^2-k^2 \vae^2)\frac{\mui}{\mue}Q_1 \\ [0.2cm]
&&
\displaystyle 
 +i \pi \rhoi \rhoe (\omega^2-k^2\vai^2)(\omega^2-k^2 \vae^2)
       \frac{G_1}{\mue}
    \left[
       \frac{k^2}{\rho_{\rm c}|\Delta_{\rm c}|}
          \left(
             \frac{c^2_{\rm sc}}{c^2_{\rm sc}+v^2_{\rm Ac}}
          \right)^2
       +\frac{1}{\rho_{\rm A}|\Delta_{\rm A}|r^2_{\rm A}}
    \right]
=0~.
 \end{array}
\end{eqnarray}
Out of the many symbols, relevant here are the definitions
    for $\Delta_{\rm c}$ and $\Delta_{\rm A}$, 
\begin{eqnarray}
\label{eq_TB2}
 \displaystyle
\Delta_{\rm c} = 
  \left.
       \frac{{\rm d}(\omega^2-k^2c_{\rm T}^2)}{{\rm d}r}\right|_{r=r_{\rm c}}~,~~ 
\Delta_{\rm A} =
  \left.
       \frac{{\rm d}(\omega^2-k^2v_{\rm A}^2)}{{\rm d}r}\right|_{r=r_{\rm A}}~,
\end{eqnarray}
    which were first introduced by \citet{1991SoPh..133..227S}. 
Here the subscript c pertains to the cusp resonance,
    which takes place at $\rc$ where 
    $\omgR = k \ct$.
Likewise, by the subscript A we refer to the \Alf\ resonance,
    which takes place at $\rA$ where $\omgR = k \va$.
We note that the terms in the square parentheses 
    derive from the jumps in the radial Lagrangian displacement ($[\tilde{\xi}]$) 
    across the resonant locations,
    and these jumps are responsible for SSKMs
        to be resonantly damped
     \citep{2011SSRv..158..289G}.
As such, the importance
    of the cusp resonance relative to the \Alf\ one
    can be measured by 
    the ratio ($X$) between the two terms, which equals 
    $[\tilde{\xi}]_{\rm c}/[\tilde{\xi}]_{\rm A}$ in the TB limit
    (see Equations~19 and 20 in Y17). 

Equation~\eqref{eq_TB} is also necessary to solve given that we would like
    to compare the damping rates in the TB limit
    with those found with the resistive approach.     
We start with an initial guess for $\omega$, thereby finding temporary values
    for $\rc$ and $\rA$.
The terms in the square parentheses are then evaluated, which enables us
    to solve Equation~\eqref{eq_TB} to update $\omega$.
With this updated $\omega$ as a further initial guess,      
    this process is iterated until convergence is met, thereby yielding
    a unique eigen-frequency that depends only on $[kR, l/R]$.
    
\subsection{Energetics of Linear Waves in Resistive MHD}
\label{sec_energetics_formulation}

It turns out necessary to examine the wave damping from the perspective of
    energy balance. 
For this purpose, we dot Equations~\eqref{eq_linMHD_momen} 
   and \eqref{eq_linMHD_Farad} 
   with $\myvect{v}_1$ and $\myvect{B}_1/\mu_0$, respectively.
Likewise, we multiply Equation~\eqref{eq_linMHD_pres} by    
   $p_1/(\gamma p_0)$.
Adding up the resulting equations then leads to
    a conservation law
\begin{eqnarray}
 \label{eq_linMHD_ener_cons}
  \frac{\partial \epsilon}{\partial t}
=
 -\nabla\cdot\myvect{f}
 -\sideal -\sres~,
\end{eqnarray}
   where
\begin{eqnarray}
\displaystyle
     \epsilon
&=&  \frac{1}{2}\rho_0\myvect{v}_1^2
    +\frac{\myvect{B}_1^2}{2\mu_0}
    +\frac{p_1^2}{2 \gamma p_0}~,
		  \label{eq_linMHD_ener_den} \\
\displaystyle
    \myvect{f}
&=&
    \ptot \myvect{v}_1
  - \frac{1}{\mu_0} (\myvect{v}_1 \cdot \myvect{B}_1) \myvect{B}_0
  + \frac{\eta}{\mu_0} \myvect{j}_1\times\myvect{B}_1
  		  \label{eq_linMHD_f_final}	\\
\displaystyle   		  
  \sideal 
&=&
  \myvect{j}_0\cdot (\myvect{v}_1\times \myvect{B}_1)
 +\frac{1}{\gamma p_0} p_1 \myvect{v}_1 \cdot \nabla p_0~,
			\label{eq_linMHD_sideal_final} \\
  \sres 
&=&
  \eta
  \left[
      \myvect{j}_1^2
    - \frac{2(\gamma-1)}{\gamma} \frac{p_1 \myvect{j}_1\cdot\myvect{j}_0}{p_0}
  \right]~.
			\label{eq_linMHD_sres_final} 		
\end{eqnarray}
Furthermore, $\ptot$ is the Eulerian perturbation
          of total pressure, 
    and $\myvect{j}_0$ ($\myvect{j}_1$) is the equilibrium (perturbed) 
    electric current density,
\begin{eqnarray}
\displaystyle 
   \ptot     = p_1 + \frac{\myvect{B}_0\cdot\myvect{B}_1}{\mu_0}~,~~~
   \myvect{j}_0 = \frac{\nabla \times \myvect{B}_0}{\mu_0}~,~~~
   \myvect{j}_1 = \frac{\nabla \times \myvect{B}_1}{\mu_0}~.
\label{eq_def_ptot_j0_j1}
\end{eqnarray} 
Integrating Equation~\eqref{eq_linMHD_ener_cons}
    over some fixed volume $V$, one finds that 
\begin{eqnarray}
\displaystyle 
  \frac{\mathd}{\mathd t} \int_{V} \epsilon {\mathd} V
=
 -\oint_{\partial V} {\mathd}\myvect{A}\cdot \myvect{f}
 -\int_{V} \left(\sideal + \sres\right) {\mathd}V~,
 			\label{eq_linMHD_ener_cons_INT}
\end{eqnarray}
    where $\partial V$ is the surface that encloses $V$.
    
Some remarks on Equation~\eqref{eq_linMHD_ener_cons} are needed. 
Firstly, within the framework of resistive MHD, 
    Equation~\eqref{eq_linMHD_ener_cons} governs the energetics of arbitrary
    small-amplitude wave-like perturbations in an arbitrary nonuniform medium,
    provided that an equilibrium flow ($\myvect{v}_0$) is absent and hence 
    the equilibrium is in force balance as governed by
\begin{eqnarray}
-\nabla p_0 (\myvect{x}) 
+ \myvect{j}_0 (\myvect{x}) \times \myvect{B}_0 (\myvect{x}) = 0~.
\label{eq_forceblance_vec}
\end{eqnarray} 
Evidently, Equation~\eqref{eq_forceblance_vec} simplifies
    to Equation~\eqref{eq_freedom} for a radially structured equilibrium.
Secondly, the ideal version ($\eta = 0$) 
    of Equation~\eqref{eq_linMHD_ener_cons} was
    first derived by \citet{1963JGR....68..147B} to our knowledge.
Well known is that $\epsilon$ and $\myvect{f}$ represent the wave energy 
    and energy flux densities, respectively. 
Less well known is the $\sideal$ term, for which \citet{1985GApFD..32..123L} 
    offered a physical interpretation where both terms on the right-hand side (RHS)
    of Equation~\eqref{eq_linMHD_sideal_final}
    are connected to the rate of
    work done on the perturbed flow field ($\myvect{v}_1$)
    by some perturbations to the \Ampere\ force density.  
The first term results directly from $\myvect{j}_0\times\myvect{B}_1$,
    while the second term derives from the perturbation to the mass density
    of a fluid parcel that experiences the unperturbed 
    \Ampere\ force
    ($\myvect{j}_0 \times \myvect{B}_0$, see
    Equation~\ref{eq_forceblance_vec}). 
Evidently, $\sideal$ is relevant, at least in principle, for wave energetics
    unless $\myvect{B}_0$ is potential. 
Thirdly, introducing a finite $\eta$ results in both the $\sres$ term
    and an additional Poyning flux
    associated with the conductive electric field
    (the last term on the RHS of Equation~\ref{eq_linMHD_f_final}). 
The first term on the RHS of Equation~\eqref{eq_linMHD_sres_final}
    evidently represents the Joule dissipation rate,
    while the second term can be loosely interpreted as resulting from
    the interaction between
    the perturbed ($\myvect{j}_1$) and unperturbed ($\myvect{j}_0$)
    electric current densities.
While a conservation law in the full form of Equation~\eqref{eq_linMHD_ener_cons}
    is not available as far as we know,
    we tend not to claim to have derived it because introducing $\eta$
    is rather straightforward. 
On top of that, energy balance can be expressed in multiple ways or
    possibly even 
    in an infinite number of ways
    \citep[see e.g.,][Section~4]{2003ApJ...599..626B}.
It suffices to note only one example, where    
    one dots Equations~\eqref{eq_linMHD_momen} and \eqref{eq_linMHD_Farad}
    with $\myvect{v}_1$ and $\myvect{B}_1/\mu_0$, respectively.
Combining the resulting equations and integrating over $V$, one ends up with
    an alternative conservation law as derived 
    in \citet[][Equation~4.3]{1989SoPh..123...83P}.   
    
Our way for examining the wave energetics differs from relevant previous ones
    in three aspects.
Firstly, we take Equation~\eqref{eq_linMHD_ener_cons}
    (or equivalently Equation \ref{eq_linMHD_ener_cons_INT}) 
    as our starting point. 
Compared with the formulation in, say, \citet{1989SoPh..123...83P},
    Equation~\eqref{eq_linMHD_ener_cons} has the advantage that $\epsilon$
    and $\myvect{f}$ have clearer physical interpretations.
For instance, involving Equation~\eqref{eq_linMHD_pres} in the derivation, 
    we have a clearly defined internal energy term in $\epsilon$.
Likewise, $\myvect{f}$ then involves explicitly the well-known acoustic flux
    $p_1 \myvect{v}_1$.
Secondly, our examination of the wave energetics is intended to expand our understanding
    of the damping rates, which are found simultaneously with the eigen-functions. 
This is different from driven problems as examined in, say, 
    \citet{1989SoPh..123...83P}, where waves with real frequencies are driven into
    the system and the rate of energy absorption in the \Alf\ continuum 
    is evaluated by working with the resistive eigen-functions
    \citep[see also][for more on driven problems]{2011SSRv..158..289G}.
Thirdly, our way is also different from the limited number of available studies
    where the philosophy is similar
    (\citeauthor{2011A&A...533A..60A}~\citeyear{2011A&A...533A..60A}, 
     \citeauthor{2013ApJ...777..158S}~\citeyear{2013ApJ...777..158S};
     see also 
     \citeauthor{1994PhPl....1..691W}~\citeyear{1994PhPl....1..691W}). 
Basically the point is that these studies were
     exclusively conducted by assuming a vanishing gas pressure
     because the focus was on the resonant absorption of
     coronal kink modes in the \Alf\ continuum.
It follows from this assumption that the equilibrium magnetic field $\myvect{B}_0$
     is potential and hence $\sideal$ is not involved.       
In our equilibrium setup, $\myvect{B}_0$ is necessarily non-potential and
     hence $\sideal$ needs to be evaluated. 
On top of that, we will examine both resonantly and resistively 
     damped modes.

\section{Numerical Results}
\label{sec_num_results}

\subsection{Overview of Different Damping Regimes}
\label{sec_num_overview}

Now we are in a position to examine the damping of SSKMs.
For this purpose we start with Figure~\ref{fig_disp_Rm},
    which shows how the dispersive properties
    of SSKMs depend on
    the magnetic Reynolds number ($\Rm$) for a number of
    combinations of $[l/R, kR]$.
Presented in Figure~\ref{fig_disp_Rm}a is the damping-time-to-period ratio 
    $\tau/P$, which derives from the real ($\omgR$, Figure~\ref{fig_disp_Rm}b)
    and imaginary ($\omgI$, Figure~\ref{fig_disp_Rm}c)
    parts of the eigen-frequency. 
The area shaded yellow pertains to the values of $\Rm$ between $10^4$
    and $6.6\times 10^7$, a range likely to apply
    to pores and sunspot umbrae.
Note that Figure~\ref{fig_disp_Rm} is a pictorial representation
    of Equation~\eqref{eq_omega_formal}, which in turn 
    represents the numerical solutions to the EVP
    in resistive MHD as laid out in Equations~\eqref{eq_Fourier_vr} to 
    \eqref{eq_Fourier_p}.
As such, the damping of SSKMs is expected
    to experience the combined effects due to Ohmic resistivity
    and resonant absorption.
Indeed, three regimes show up for all pairs of $[l/R, kR]$.
When $\Rm$ is sufficiently large, all curves in Figure~\ref{fig_disp_Rm}a 
    level off, 
    a defining signature for resonant absorption to be solely
    responsible for damping collective modes
    \citep{1991PhRvL..66.2871P}.
We therefore label this regime as the ``resonant regime".    
Moving away from this regime toward smaller $\Rm$, one sees that 
    $\tau/P$ tends to decrease until a knee is reached, beyond which
    $\tau/P$ possesses a simple linear $\Rm$-dependence.    
We call this last regime the ``Ohmic regime" because the damping of SSKMs 
    can be solely attributed to Ohmic resistivity. 
To show this, we note that $\tau/P \propto \omgR/|\omgI|$.
We note further that
    the axial phase speeds ($\vph = \omgR/k$, Figure~\ref{fig_disp_Rm}b)
    of SSKMs differ little from $\cti$ for all combinations of $[l/R, kR, \Rm]$
    examined in this study, meaning in particular
    that $\vph$ and hence $\omgR$ are essentially constants for a given $[l/R, kR]$
    in the present context.
On the other hand, for a given $[l/R, kR]$, 
    one intuitively expects that $|\omgI| \propto \eta \propto 1/\Rm$ for
    weak damping, 
    which is indeed the case as demonstrated by Figure~\ref{fig_disp_Rm}c.
The end result is then $\tau/P \propto \Rm$.
The range of $\Rm$ that lies between the two regimes is to be 
    named the ``intermediate regime'', where Ohmic resistivity
    and resonant absorption both play a role for damping SSKMs.
This intermediate regime is not as clearly separated from 
    the resonant regime as it is from the Ohmic one.
For definitiveness, we nonetheless regard the resonant regime
    to start at an $\Rm$ beyond which
    $\tau/P$ varies by no more than $1.5\%$ over a decade or so.
    
Overall, the point to draw from Figure~\ref{fig_disp_Rm} is that 
	in general the importance of Ohmic resistivity relative to resonant absorption
	for damping SSKMs
	needs to be assessed on a case-by-case basis, given the broad coverage
	of the shaded area.  
That said, the damping rates can be order- or even orders-of-magnitude
    larger than the ones deriving from 
    resonant absorption when Ohmic resistivity dominates, as happens for waveguides
    with, say, small sizes and/or weak magnetic fields (see the definition for $\Rm$).
This is true despite that $\Rm$ consistently exceeds $10^4$, a value that seems
    to be sufficiently large for one to propose resonant absorption
    as the primary damping mechanism by drawing experience from pertinent
    coronal studies.
Indeed, when coronal radial fundamental kink modes are of interest, 
    the \Alf\ resonance tends to become the sole
    damping mechanism when $\Rm \gtrsim 3 \times 10^3 - 10^5$
    (see e.g., 
    Figure~2 in \citeauthor{2006ApJ...642..533T}~\citeyear{2006ApJ...642..533T};
    Figure~9 in \citeauthor{2016SoPh..291..877G}~\citeyear{2016SoPh..291..877G}). 
These values are many orders-of-magnitude smaller than typically accepted
    coronal values for $\Rm$, making resonant absorption a leading mechanism
    to account for the damping of the abundantly measured coronal kink modes
    \citep{1992SoPh..138..233G,2002ApJ...577..475R,2002A&A...394L..39G}.
What Figure~\ref{fig_disp_Rm} highlights is that
    in general Ohmic resistivity needs to be incorporated
    in studies of collective modes in photospheric waveguides.
This point has already been raised by both C18 and \citet{2020ApJ...897..120G}
    who examined photospheric SSSMs,
    and derives from the considerable difference between the physical conditions
    of photospheric waveguides and the coronal ones.
As an example, we note that the waveguides
   in this study are under-dense ($\rhoi/\rhoe<1$, Figure~\ref{fig_EQprofile}b),
        whereas bright coronal loops are consistently over-dense ($\rhoi/\rhoe > 1$).

The $\Rm$-dependence of the behavior of SSKMs
    can also be examined from the perspective of energy balance.
We start by rewriting the Fourier ansatz
    (Equation~\ref{eq_Fourier}) as
\begin{eqnarray*}
   g_1 (r, \theta, z; t)
 = {\rm Re}\{\tilde{g}(r) \exp[-i(\omgR  t- kz -\theta)]\} \exp(\omgI t)~,
\end{eqnarray*}    
    such that the oscillatory ($\omgR$)
    and damping ($\omgI$) time-dependencies are explicitly separated. 
One then finds that the product of any two first-order quantities $g_1$ and $h_1$, 
    when averaged over one axial wavelength ($\lambda = 2\pi/k$), 
    evaluates to
\begin{eqnarray}
  \left< g_1(r, \theta, z; t) h_1(r, \theta, z; t) \right>
\equiv \frac{1}{\lambda} \int_{0}^{\lambda} 
      g_1(r, \theta, z; t) h_1(r, \theta, z; t) {\mathd} z 
=  \overline{g_1 h_1} \exp(2\omgI t)~,   
  			\label{eq_def_bracket}
\end{eqnarray}
    where
\begin{eqnarray}
  \overline{g_1 h_1} 
 =\frac{1}{2}{\rm Re}\left\{\tilde{g}^*(r)\tilde{h}(r)\right\}    
 =\frac{1}{2}{\rm Re}\left\{\tilde{g}(r)  \tilde{h}^*(r)\right\}~.    
 			\label{eq_def_bar}
\end{eqnarray}
The asterisks represent the complex conjugate, and  
     $\overline{g_1 h_1}$ is seen to depend only on $r$.
Now take a volume $V$ to be the space bounded by two transverse planes
     that are an axial wavelength apart. 
One readily finds that the net wave energy flux leaving this volume vanishes,
     given that only trapped modes are of interest
     (see the surface integral in Equation~\ref{eq_linMHD_ener_cons_INT}).
After some further algebra, Equation~\eqref{eq_linMHD_ener_cons_INT} becomes
\begin{eqnarray}
  -2\omgI E
= S_{\rm ideal} + S_{\rm res}~,
   \label{eq_linMHD_ener_cons_Eig_EntireVol}
\end{eqnarray}
    where
\begin{eqnarray}
\displaystyle 
  E     = \int_0^\infty \bar{\epsilon}(r')       r'{\mathd}r'~,~~ 
\Sideal = \int_0^\infty \bar{s}_{\rm ideal} (r') r'{\mathd}r'~,~~
\Sres   = \int_0^\infty \bar{s}_{\rm res} (r')   r'{\mathd}r'~.
        \label{eq_def_E_S_inf} 
\end{eqnarray} 
The barred quantities in the integrand can be readily evaluated by applying
    Equation~\eqref{eq_def_bar} to the products of first-order quantities
    involved in Equations~\eqref{eq_linMHD_ener_den} 
    to \eqref{eq_linMHD_sres_final}.    
Physically speaking, Equation~\eqref{eq_linMHD_ener_cons_Eig_EntireVol} 
    means that the wave damping is realized in two channels, 
    one through the Ohmic resistivity and the other through the mutual
    interaction between the waves and the equilibria.

Figure~\ref{fig_enerbal_Rm} examines the $\Rm$-dependence of some key
    terms in Equation~\eqref{eq_linMHD_ener_cons_Eig_EntireVol} 
    for the same set of combinations $[l/R, kR]$ as
    in Figure~\ref{fig_disp_Rm}. 
The shaded area is also inherited from there.     
Let us start with Figure~\ref{fig_enerbal_Rm}a, where
    $(\Sideal+\Sres)/E$ is plotted in conjunction with $-2 \omgI$ 
    (the black diamonds).
Note that the latter is essentially a direct output from our code,
    whereas the former needs to be evaluated with the eigen-functions.     
With Equation~\eqref{eq_linMHD_ener_cons_Eig_EntireVol}, one expects that
    the diamonds should lie on the curves, which is seen to be indeed the case. 
In fact, a close agreement between $-2\omgI$
    and $(\Sideal+\Sres)/E$ is found for all values of $\Rm$ at a
    given $[l/R, kR]$, with the diamonds representing only a small subset
    in order not to obscure the curves. 
On the one hand, this agreement validates the remarkable numerical accuracy of
    our computations.
On the other hand, the curves leave out the explicit $k$-dependence
    and hence are less crowded than in Figure~\ref{fig_disp_Rm}c,
    making the three regimes of wave damping more evident.          
Now move on to Figure~\ref{fig_enerbal_Rm}b, where $\Sideal/\Sres$ 
    is plotted to evaluate the relative importance of the two terms.
This evaluation is informative in that as far as we know, 
    the $\Sideal$ term has not been explicitly
    examined, at least not in the context of photospheric modes.
To proceed, we note that the $\Sres$ term is always positive
    for all the computations because it is
    by far dominated by
    the Joule dissipation rate (see Equation~\ref{eq_linMHD_sres_final}). 
In contrast, the $\Sideal$ term can be either positive or negative, thereby 
    serving as a sink or source in physical terms.    
One further sees that $\Sres$ dominates $\Sideal$ when $\Rm$
    is either sufficiently large or sufficiently small, or equivalently
    when the modes are very deep in either the Ohmic or the resonant regime.
When $\Sideal/\Sres$ is appreciable, one sees that its magnitude tends to be stronger
    for smaller values of $l/R$ at a given $kR$
    (see the red curves with different linestyles).
Likewise, this magnitude tends to be stronger for smaller $kR$
    when $l/R$ is given (see the solid curves with different colors).
These details aside, we conclude that $\Sideal$ needs
    to be considered for examining the energetics of SSKMs in our model equilibria,
    given that $|\Sideal/\Sres|$ may reach $\sim 28\%$ 
    in the case where $[l/R, kR] = [0.1, 0.7]$.  

Some further understanding of the different damping regimes
    can be gained from Figure~\ref{fig_Sres_r_distri}, where
    $\bar{s}_{\rm res}$ is plotted as a function of $r$ for
    a fixed pair of $[l/R, kR] = [0.1, 0.7]$.
Here the shaded area represents the transition layer (TL).      
A number of values for $\Rm$ are adopted, and the corresponding 
    $\bar{s}_{\rm res}$ is rescaled such that its maximum attains unity. 
Above all, one sees that $\bar{s}_{\rm res}$ almost vanishes identically
    in the exterior. 
The black curve in Figure~\ref{fig_Sres_r_distri}a pertains to 
    $\lg\Rm = 4.3$, and is representative of the Ohmic regime.
Most notable in this case is that 
    $\bar{s}_{\rm res}$ is extended over both the interior
    and the TL. 
On the contrary, 
    pertaining to an $\Rm$ fairly deep in the resonant regime,
    the red curve in Figure~\ref{fig_Sres_r_distri}b is characterized
    by the concentration of $\bar{s}_{\rm res}$
    in two very narrow layers. 
Evidently, the inner and outer layers correspond to the cusp
    and \Alf\ resonances, respectively.
Given that the axial phase speeds of resonantly damped SSKMs
    are close to the internal tube speed 
    ($\cti$, see Figure~\ref{fig_disp_Rm}b),     
    one would have expected the locations of the resonances 
    in view of the radial profiles of $\ct$ and $\va$ (Figure~\ref{fig_EQprofile}). 
Examining the rest of the curves in Figure~\ref{fig_Sres_r_distri},
    one sees that $\bar{s}_{\rm res}$ becomes increasingly concentrated
    around the resonances as $\Rm$ increases.
This behavior also offers a qualitative explanation for the rather involved behavior
    of $\Sideal/\Sres$ shown in Figure~\ref{fig_enerbal_Rm}b. 
For this purpose we note that by definition, $\Sideal$ ($\Sres$) 
    collects all contributions from $\sideal$ ($\sres$) over
    the entire range of $r$.
However, $\sideal$ does not vanish only in the TL. 
If then follows that $\Sideal/\Sres$ is determined by two factors,
    one being how wide $\sideal$ is distributed relative to $\sres$
    and the other being how $\sideal$ compares with $\sres$ at specific locations.
It turns out that at sufficiently small $\Rm$, the former factor is more important,
    leading to small values of $\Sideal/\Sres$ despite that $\sideal$ can be comparable
    to $\sres$ locally.
When $\Rm$ is sufficiently large, the latter factor is more important,
    once again resulting in small values of $\Sideal/\Sres$ even though
    $\sideal$ can be distributed in a more extended manner.      
  
One may have noticed two peculiarities in Figure~\ref{fig_disp_Rm}a, 
    one being that some curves do not
    extend to the largest $\Rm$ in the plot, the other being that
    some curves are absent for some combinations of $[l/R, kR]$.
Both result from some limitations of our numerical method.
Recall that, given a pair of $[l/R, kR]$,
    we first look for the resonant regime and 
    then vary $\Rm$ toward the target value such that a unique eigen-frequency
    can be found. 
Two technical issues arise.
When $\Rm$ is extremely large,
    sometimes the computational resources
    required for resolving the extremely fine scales where
    resonant absorption occurs may be too demanding.
This explains the first peculiarity. 
Another issue is that 
    sometimes our code does not converge to a unique solution when we reduce $\Rm$ 
    from the resonant regime. 
This happens when $kR$ is large for a given $l/R$ or when 
    $l/R$ is large for a given $kR$. 
Note that this is not to say that a proper resonant regime cannot be found
    in these situations.
It is just that for a given pair of $[l/R, kR]$, 
    the $\tau/P$ profile spans only a narrow range of $\Rm$
    such that it is not straightforward
    to address the importance of 
    the Ohmic resistivity for damping the SSKMs. 
These technical issues notwithstanding, 
    we are allowed to further examine resonant absorption by capitalizing
    on the computability of the resonant regime for those pairs of $[l/R, kR]$
    that cover a much broader range than presented in Figure~\ref{fig_disp_Rm}. 
This examination is important in its own right, because resonant absorption
    is indeed the sole damping mechanism 
    when $\Rm$ is sufficiently large.

\subsection{Resonantly Damped SSKMs}
\label{sec_num_reso}
Figure~\ref{fig_disp_lR} presents the $l/R$-dependence of
        (a) the damping-time-to-period-ratios ($\tau/P$),
        (b) the axial phase speeds ($\omgR/k$),
    and (c) the damping rates ($\omgI$) in units of $k\vai$ 
   for resonantly damped SSKMs for a number of axial wavenumbers ($kR$).
In addition to the values computed with resistive MHD (the solid curves),
    Figure~\ref{fig_disp_lR} also presents the TB expectations (dashed) found by
    solving Equation~\eqref{eq_TB}.
On top of that, we have also plotted the results found by solving Equation~\eqref{eq_TB}
    with the second term in the square parentheses neglected (the dotted curves).
For the ease of description, 
    let ``full TB'' and ``cusp TB''
    refer to the dashed and dotted curves, respectively.
One may question why the ``cusp TB'' results are shown 
    together with the ``full TB'' ones.
We will address this question shortly, because it seems better to start with
    a comparison between the resistive and ``full TB'' results.
Examining Figure~\ref{fig_disp_lR}a, 
    one sees that the smaller the value of $l/R$, 
    the smaller the difference between the solid and dashed curves,
    a qualitative behavior implied in the TB approximation by construction.
What needs to be examined is how small an $l/R$ should be for the ``full TB'' results
    to well agree with the resistive results for a given $kR$.  
We quantify this by first defining 
    $\epsilon \equiv |(\tau/P)^{\rm res}/(\tau/P)^{\rm TB}-1|$
    to measure the difference in the values for $\tau/P$ found with the two approaches. 
We further define $(l/R)^{\rm TB}$ to be the critical value only below which
    $\epsilon \le 30\%$.
One finds that in general $(l/R)^{\rm TB}$ decreases with $kR$, reading
    $0.074$ ($0.026$) for a $kR$ of $0.3$ ($4.3$).
This $k$-dependence of $(l/R)^{\rm TB}$ was also found for photospheric SSSMs in C18,
    and can be explained by the same intuitive reasoning therein.
After all, with the axial wavelength ($\lambda = 2\pi/k$) also a relevant lengthscale, 
    the TL width ($l$) should be much smaller than both $R$ and $\lambda$ for 
    the TB limit to hold.
Somehow surprising is that $(l/R)^{\rm TB}$ is so small.
Indeed, $\epsilon$ evaluates to $2.6$ for an $l/R$ as small as $0.1$ when 
    $kR = 4.3$.  
For this pair of $[l/R, kR]$, one sees
    that the substantial value for $\epsilon$
    results from the differences in the $l/R$-dependence 
    of the resistive and ``full TB'' profiles. 
While the $\tau/P$ profile in the TB limit 
    decreases monotonically with $l/R$ in the plotted range,
    its resistive counterpart possesses a non-monotonic behavior as characterized
    by the appearance of a minimal $\tau/P$.
On the other hand, one sees from Figure~\ref{fig_disp_lR}b
   that the axial phase speeds ($\omgR/k$) are consistently
   smaller than $\cti$ by only a small amount.
It then follows that the behavior of $\tau/P$, 
   the existence of a minimal $\tau/P$ in particular,
   derives from the $l/R$-dependence of the 
   imaginary part of the eigen-frequencies 
   ($\omgI$, Figure~\ref{fig_disp_lR}c).   
We note that
    the existence of a minimal $\tau/P$ holds for all the resistive
    computations we have conducted.     
We note {further} that this non-monotonic $l/R$-dependence has been seen
    for the resonantly damped
    photospheric SSSMs (C18) and
    coronal radial fundamental kink modes alike
    \citep[][S13 hereafter]{2013ApJ...777..158S}.
This S13 study approached the resonant absorption of coronal kink modes
    in the \Alf\ continuum from the ideal quasi-mode perspective with 
    the Frobenius method.
Furthermore, the non-monotonic $l/R$-dependence of the damping rates for a
    particular transverse profile (Figure~6 in S13) was interpreted from 
    energetics considerations by evaluating
    the energy flux injected into the resonance and the integrated wave energy.
In what follows, we analyze the $l/R$-dependence of the damping rates
    of SSKMs from the energetics perspective as well.

We start by considering the entire volume bounded by two transverse planes that are
    separated by an axial wavelength. 
We then exclude two cylindrical shells, the inner shell
    bounded by $\rcminus$ and $\rcplus$, and the outer one
    bounded by $\rAminus$ and $\rAplus$. 
Let $V$ denote the resulting volume.    
The subscripts ${\rm c}$ and ${\rm A}$ are adopted because these shells
    will be associated with the cusp and \Alf\ resonances shortly.
For now it suffices to assume that $\rcminus < \rcplus < \rAminus <\rAplus$,
    and that
    the Ohmic resistivity is negligible for determining
    the eigen-functions in $V$.         
Equation~\eqref{eq_linMHD_ener_cons_INT} then becomes
\begin{eqnarray}
  -2\omgI \hat{E}
=  \hatFc + \hatFA + \hat{S}_{\rm ideal}~,
   \label{eq_linMHD_ener_cons_Eig_shell}
\end{eqnarray}
    where
\begin{eqnarray}
\displaystyle 
\hat{E}  &=& 
   \left(
     \int_0^{\rcminus} + \int_{\rcplus}^{\rAminus} + \int_{\rAplus}^\infty
   \right)\bar{\epsilon}(r') r'{\mathd}r'~, 
   				\label{eq_def_hatE} \\
\hat{S}_{\rm ideal}  &=& 
   \left(
     \int_0^{\rcminus} + \int_{\rcplus}^{\rAminus} + \int_{\rAplus}^\infty
   \right)\bar{s}_{\rm ideal}(r') r'{\mathd}r'~, 
   				\label{eq_def_hatSideal} \\[0.1cm]
\hatFc &=& 
   \rcminus \bar{f}_r(\rcminus) - \rcplus \bar{f}_r(\rcplus)
      \equiv \left\{ r \bar{f}_r \right\}_{\rm c}~,
   				\label{eq_def_Fc} \\[0.1cm]
\hatFA &=& 
   \rAminus \bar{f}_r(\rAminus) - \rAplus \bar{f}_r(\rAplus)
       \equiv \left\{ r \bar{f}_r \right\}_{\rm A}~.
   				\label{eq_def_FA} 
\end{eqnarray} 
Note that we have introduced the brace operator $\{q\}$
    to evaluate the difference of some quantity $q$ across a shell.
In addition, $\bar{f}_r$ is found by applying the bar operation
    (Equation~\ref{eq_def_bar}) to the radial component of 
    the wave energy flux density 
    ($\myvect{f}$, Equation~\ref{eq_linMHD_f_final}).  
In the absence of the Ohmic resistivity, $f_r$ simplifies to $\ptot v_{1r}$
    and consequently $\bar{f}_r = {\rm Re}(\ptottilde^* \tilde{v}_r)/2$.
Evidently, Equation~\eqref{eq_linMHD_ener_cons_Eig_shell} means
    that the wave energy in the volume $V$ is lost via both
    net fluxes into the shells ($\hatFc$ and $\hatFA$)
    and the interactions with the equilibrium ($\hat{S}_{\rm ideal}$). 
     
With Equation~\eqref{eq_linMHD_ener_cons_Eig_shell} we recall
   some features of the ideal MHD quasi-mode approach such that some intricacies
   with our resistive MHD approach can be appreciated.
As far as resonantly damped modes are concerned,
   the ideal quasi-mode approach has been exclusively conducted
   by assuming a vanishing gas pressure 
   (S13, and \citeauthor{2017ApJ...850..114S}~\citeyear{2017ApJ...850..114S})
   \footnote{Here by ``ideal" we specifically refer to the quasi-mode approach
   where dissipative effects are discarded from the outset. 
   To our knowledge, this approach has been realized only through
       the Frobenius method formulated in 
       \citet{1994PhPl....1..691W} and first practiced in the solar context
       by S13. 
   Some approaches can be regarded as nearly ideal, with the formulation developed
       in \citet{1991SoPh..133..227S} as a much-employed prototype.
   Restricting ourselves to Equation~\eqref{eq_TB}, we mean by ``nearly ideal"
       that the outcomes from these approaches do not explicitly involve
       dissipative effects but dissipative MHD is nonetheless involved
       in the mathematical procedure.  
   }.
It then follows that only the \Alf\ resonance arises,
   and $\sideal$ vanishes identically. 
On top of that, the Ohmic resistivity is conceptually irrelevant
   in the first place, and resonant absorption can be visualized as taking
   place in a resonant layer that is infinitely thin ($\rAminus = \rAplus$).
In view of Equation~\eqref{eq_linMHD_ener_cons_Eig_shell}, that $\omgI$ does not vanish
   in S13 is then due to the discontinuity
   in $\bar{f}_r$ across the \Alf\ resonance, which in turn derives from the discontinuities in $\ptottilde$ and $\tilde{v}_r$ (see Figure~3 therein). 
Without concrete ideal MHD computations, we refrain from discussing further
   how resonantly damped SSKMs behave when they are eventually found this way.  
Rather, it suffices to note that the ideal eigen-frequencies are expected to agree
   with their resistive counterparts but this is not true regarding the
   eigen-functions. 
This latter aspect has some bearings on how we evaluate the terms
   in Equation~\eqref{eq_linMHD_ener_cons_Eig_shell} with our resistive eigen-functions.
The point is, if neglecting the subscripts ${\rm c}$ and ${\rm A}$,
   then one finds that Equation~\eqref{eq_linMHD_ener_cons_Eig_shell}
   holds for an arbitrary $V$ as long as resistivity can be neglected therein.   
Evidently, we need to associate the two cylindrical shells with the resonances for the
   evaluation to make physical sense. 
However, some intricacies arise regarding how to determine the borders of 
   the shells, known as dissipative layers
   \citep[DLs, e.g.,][]{2011SSRv..158..289G}.     
For the much-studied \Alf\ resonance, the DL width ($\delta_{\rm A}$)
   is known to take the larger
   one out of the resistive ($\delta_{\rm A, res}$) and
   non-stationarity scales ($\delta_{\rm A, NS}$).
Here $\delta_{\rm A, res}$ depends on
   the Ohmic resistivity as $\eta^{1/3}$ \citep{1991SoPh..133..227S},
   whereas $\delta_{\rm A, NS}$ is independent of $\eta$
   but proportional to $|\omgI|$ \citep{1995JPlPh..54..129R,1996ApJ...471..501T}.
Regarding the DL width pertaining to the cusp resonance ($\delta_{\rm c}$), 
   a resistive
   spatial scale ($\delta_{\rm c, res}$) is also relevant
   and possesses the same $\eta$-dependence      
   \citep{1991SoPh..133..227S}.
Likewise, the cusp version ($\delta_{\rm c, NS}$)
   of the non-stationarity scale 
   was shown to be formally the same as $\delta_{\rm A, NS}$ 
   \citep[][see the discussions in connection to Equations~17 to 19 therein]{1998ApJ...503..422T}. 
It then follows that
   the DL widths ($\delta_{\rm c}$ and $\delta_{\rm A}$) in the present context
   will be $\Rm$-independent when $\Rm$ becomes sufficiently large.
In this case, no $\Rm$-dependence will show up for the hatted terms
   in Equation~\eqref{eq_linMHD_ener_cons_Eig_shell},
   making their evaluation more definitive. 
However, our code does not allow $\Rm$ to increase indefinitely. 
This means in particular that $\delta_{\rm A}$ never reaches the non-stationarity
   scale in our computations, and the hatted terms in
   Equation~\eqref{eq_linMHD_ener_cons_Eig_shell} are in general 
   dependent on $\Rm$.  
   
So how do we evaluate the hatted terms, 
   supposing a fixed pair of $[l/R, kR]$?
One possible way is to draw experience from the analytical study 
   by \citet{1995SoPh..157...75G}, who showed that 
\begin{eqnarray}
   r_{\rm A}^{\pm} \approx \rA \pm 5 \delta_{\rm A}~,
   \label{eq_rApm_G95}
\end{eqnarray}   
   for the \Alf\ resonance,
   provided that $\delta_{\rm A}$ is determined by $\delta_{\rm A, res}$.
However, Equation~\eqref{eq_rApm_G95} is no longer valid
   when $\delta_{\rm A}$ is determined
   by $\delta_{\rm A, NS}$ 
   \citep[e.g.,][Figure~1]{1996ApJ...471..501T}.
We therefore choose an empirical procedure to determine
   $r_{\rm A, c}^{\pm}$ for a given $\Rm$.
The idea is simply that by definition, resistivity is important only
   in the DLs, namely the two shells in the present context.
Basically, we compare the radial profile of some eigen-function against
   that obtained at a substantially larger $\Rm' = \zeta_{R} \Rm$, and locate the shells
   by looking for where the relative difference between the two profiles 
   exceeds some critical value $\zeta_{\rm cri}$. 
We consistently choose $|\tilde{v}_r|$ as the eigen-function, 
   and adopt a $\zeta_{R} = 2$.  
In addition, we take $\zeta_{\rm cri}$ to be $4 \times 10^{-4}$ 
   and $10^{-5}$ to determine the cusp and \Alf\ DLs, respectively. 
Employing other reasonable values for $\zeta_{R}$ or $\zeta_{\rm cri}$
   leads only to insignificant differences to our results.       
Figure~\ref{fig_reson_Revr_r} displays the radial profiles
   of ${\rm Re}\tilde{v}_r$ for a number of $\Rm$ 
   as labeled, with $[l/R, kR]$ fixed at $[0.2, 2]$. 
Here two different intervals are distinguished to show the details
   in the DLs pertaining to the cusp (Figure~\ref{fig_reson_Revr_r}a)
   and \Alf\ (Figure~\ref{fig_reson_Revr_r}b) 
   resonances.   
Each profile is rescaled such that $\ptottilde = 1$ at $r = 1.25~R$, otherwise 
   the comparison between different profiles will not make sense. 
Note that the details for rescaling the eigen-functions are not essential.
Rather, with Figure~\ref{fig_reson_Revr_r}a we show that 
   the width of the cusp DL ($\delta_{\rm c}$) decreases when $\Rm$ increases,
   accompanied by the appearance of more and more oscillations. 
This behavior is in line with what was pointed out by \citet{1995JPlPh..54..129R}
   and numerically demonstrated by S13 for coronal kink modes resonantly absorbed
   {in the \Alf\ continuum
   \citep[see also][Figure~2.3]{2003phd..Leuven..Andries03}.
If $\Rm$ further increases, 
   both studies showed that the width of the \Alf\ DL ($\delta_{\rm A}$) 
   will eventually be determined by $\delta_{\rm A, NS}$.
In other words, eventually the \Alf\ DL width will become a constant,
   and the effect of increasing $\Rm$ is to introduce an increasing number
   of oscillations into the DL. 
We can discern some similar signature for the cusp resonance here in that
   $\delta_{\rm c}$ is found to depend on $\Rm$ only weakly when $\lg\Rm \gtrsim 9$.
With Figure~\ref{fig_reson_Revr_r}b as an illustration, 
   we note that the \Alf\ DLs in all of our computations 
   consistently narrow when $\Rm$ increases, showing 
   no evidence that non-stationarity ultimately decides $\delta_{\rm A}$
   for the SSKMs in the $\Rm$-range that we examine.

Some remarks are necessary in this context.
Our first set of remarks is connected with Figure~\ref{fig_reson_Revr_r}, 
    from which one sees that the oscillations in the DL(s) 
    can be very strong when $\Rm$ is large.
This does not invalidate the examination of resonantly damped modes
    from the perspective of dissipative eigen-modes.
Rather, this means that if resonant absorption is examined from
    the initial-value-problem (IVP) perspective, then the perturbation energy is
    increasingly transferred from global coherent motions to localized motions
    around resonant surfaces.
For coronal radial fundamental kink modes, this was established analytically
    by \citet{2002ApJ...577..475R} and has been shown by a considerable number
    of numerical studies 
    \citep[e.g.,][to name only a few]{2006ApJ...642..533T,               2015ApJ...803...43S,2020ApJ...893..157E,2020ApJ...904..116G}.
To our knowledge, however, a similar study on photospheric modes
    from the IVP perspective has yet to appear.
Our second set of remarks pertains to Figure~\ref{fig_disp_Rm}, 
    where the damping of SSKMs is seen to undergo a continuous transition
    from the Ohmic to the resonant regime.
We note that a similar problem was addressed for surface \Alf\ waves 
    in a one-dimensional slab equilibrium by \citet{1996JPlPh..56..107R}, 
    even though viscosity was adopted as the dissipative factor therein.
The transition was shown to be organized by a dimensionless
    parameter ($R_{\rm g}$) 
    that involves the relative variation of the local \Alf\ speed,
    the ratio of the TL width to the axial wavelength,
    and the viscous Reynolds number.
Our Figure~\ref{fig_disp_Rm} plots the damping rates 
    as a function of $\Rm$ for a given pair of $[l/R, kR]$, showing that 
    different damping regimes can indeed be told apart. 
However, the range of $\Rm$ pertaining to, say, the intermediate regime 
    is different for different combinations of $l/R$ and $kR$.
If some $R_{\rm g}$ can be established for photospheric modes
    in light of \citet{1996JPlPh..56..107R}, then one expects a unified range of
    $R_{\rm g}$ for the intermediate regime for different computations,
    leading to a substantially simpler way for discriminating different damping regimes.

We are finally ready to examine the $l/R$-dependence of 
   the damping rates with the aid of Equation~\eqref{eq_linMHD_ener_cons_Eig_shell}.
It suffices to examine a fixed axial wavenumber of $kR = 2$.
Figure~\ref{fig_reson_damp_l_energetics} presents, with circles of different colors,
   the $l/R$-dependencies of $\hatFA/\hat{E}$,
   $\hatFc/\hat{E}$, and $\hat{S}_{\rm ideal}/\hat{E}$,
   as well as the sum of the three ratios
   for a number of $\Rm$.
Note that the range of $\Rm$ in general depends on $l/R$.
This is due to two reasons, one being that   
   SSKMs at different $l/R$ in general
   enter into the resonant regime at different $\Rm$, 
   and the other being the technical difficulty for us
   to conduct computations when $\Rm$ is too large. 
With the exception of the sum, the symbols representing a given quantity 
   for a given $\Rm$ are connected by a solid line with the same color coding.
At any examined $l/R$, one sees that the circles of different colors 
   representing the sum cannot be distinguished from one another.
This is understandable because they evaluate to $-2\omgI$, 
   which is $\Rm$-independent.
In fact, replotting the pertinent set of values for $\omgI$ with
   the dashed curve, we see the expected behavior for the sums to agree
   with $-2\omgI$.
On the other hand, one sees that $\hatFA/\hat{E}$ increases monotonically with $l/R$
   for all the examined values of $\Rm$.
Put another way, the wave damping due to the \Alf\ resonance is stronger
   for a larger $l/R$.
Two reasons are therefore responsible for 
   the apparently counter-intuitive behavior for $|\omgI|$
   to decrease with $l/R$ when $l/R \gtrsim 0.12$.
Primarily, it happens because 
   the wave damping due to the cusp resonance behaves this way. 
To a lesser but non-negligible extent, it happens because
   $\hat{S}_{\rm ideal}/\hat{E}$ consistently weakens with $l/R$ 
   in this range of $l/R$.

An important product of Figure~\ref{fig_reson_damp_l_energetics}
   is that $\hatFc$ consistently exceeds $\hatFA$, meaning that
   the cusp resonance is consistently more important than the \Alf\ one
   for damping the SSKMs with this particular $kR$.
What about the relative importance for other values of $kR$?
Going back to Figure~\ref{fig_disp_lR}, one can most readily evaluate
   this            
   for the TB results, given the availability of
     an explicit dispersion relation (Equation~\ref{eq_TB}). 
In this case, one is allowed to isolate, say, the cusp resonance by
     keeping only the first term in the square parentheses, as we have done
     when producing the ``cusp TB'' curves.
Recall that the ratio of the first to the second term is denoted by $X$.
To proceed, 
     let $\omgI^{\rm c}$ and $\omgI^{\rm A}$ 
     denote the damping rates due to the cusp and \Alf\ resonances alone, 
     respectively. 
Likewise, let $\omgI^{\rm TB}$ denote the damping rate when both resonances 
     are accounted for. 
It then follows that $\omgI^{\rm TB} \approx \omgI^{\rm c} + \omgI^{\rm A}$
     and $\omgI^{\rm c}/\omgI^{\rm A} \approx X$,
     when the damping is sufficiently weak.
Now let $Y$ denote the ratio of the ``cusp TB'' value for $\tau/P$ 
	 to the ``full TB'' one.      
It further follows that $Y \approx 1+1/X$ because $\tau/P \propto \omgR/|\omgI|$. 
In other words, one expects that $Y < 2$ when the cusp resonance
     is stronger ($X>1$) and $Y \approx 1$ when it dominates ($X \gg 1$). 
With this preparation, let us examine Figure~\ref{fig_disp_lR}a and focus on 
     where the TB limit holds ($l/R \le (l/R)^{\rm TB}$). 
An inspection of the green curves pertinent to $kR = 0.7$ 
     indicates that $Y$ happens to be $\approx 2$, corresponding to the situation
     where the two resonances are equally important.
The cusp (\Alf) resonance becomes increasingly important when 
     $kR$ increases (decreases) from this critical value.      
Note that the transition of the relative importance of the resonances
     consistently takes place at $kR \approx 0.7$
     when $l/R \le (l/R)^{\rm TB}$.
For now it suffices to consider a given $l/R$.      
The existence of a critical $kR$ is not surprising in view of
    Equation \eqref{eq_TB}, which
    suggests that
\begin{eqnarray}
\label{eq_X_TB}
X = k^2 \Lambda~,
\end{eqnarray}
    where
\begin{eqnarray}
\label{eq_Lambda}
\displaystyle 
\Lambda =  \rA^2 
     \frac{\rho_{\rm A}}{\rho_{\rm c}}
     \frac{|\Delta_{\rm A}|}{|\Delta_{\rm c}|}  
     \left(\frac{c^2_{\rm sc}}{c^2_{\rm sc}+v^2_{\rm Ac}}
          \right)^2
  = \rA^2 
       \frac{\rho_{\rm A}}{\rho_{\rm c}}
       \frac{|\mathd \va^2/\mathd r|_{r = \rA}}
            {|\mathd \ct^2/\mathd r|_{r = \rc}}  
       \left(\frac{c^2_{\rm sc}}{c^2_{\rm sc}+v^2_{\rm Ac}}
            \right)^2~.
\end{eqnarray}     
We arrive at the second equality by 
   replacing $\Delta_{\rm A}$ and $\Delta_{\rm c}$
   with Equation~\eqref{eq_TB2}.
If one assumes that $\rA \approx \rc \approx R$, 
   then Equation~\eqref{eq_X_TB} simplifies to 
   Equation~(14) in \citet{2009ApJ...695L.166S}, where 
   the effect of the cusp continuum was examined
   for damping the transverse oscillations in prominence threads. 
Given the appearance of $k^2$ in $X$, one would expect 
   the cusp resonance to dominate for large $kR$, an expectation that
   indeed holds.  
Less trivial is at which $kR$ 
   the cusp resonance starts to dominate.     
When $k$ varies, it turns out that the $k$-dependence of $\Lambda$
   derives essentially from the $k$-dependence 
   of $(\mathd \ct^2/\mathd r)_{r = \rc}$.
This subtlety results from the extremely weak dispersion 
   of SSKMs (see Figure~\ref{fig_disp_lR}b).
With $\omgR/k$ close to $\cti$, neither $\rc$ nor $\rA$
   is sensitive to $k$, and hence the insensitivity     
   of nearly all the terms involved in $\Lambda$,  
   with $(\mathd \ct^2/\mathd r)_{r = \rc}$ being the only exception. 
However weak its $k$-dependence is,    
   $\omgR/k$ nonetheless increases toward $\cti$ with decreasing $k$.
Therefore the cusp resonance point ($\rc$) moves toward the inner
   boundary of the transition layer (see Figure~\ref{fig_EQprofile}a),
     thereby leading to 
   some substantial decrease in $|\mathd \ct^2/\mathd r|_{r = \rc}$
   (see Equation~\ref{eq_profile_cs_ct})
\footnote{
   It can be readily shown 
       that $\mathd \ct^2/\mathd r \propto \delta r \equiv (r-r_{\rm i})$ for a given $l/R$
       when $0 < |\delta r| \ll l$.    
   Take $l/R = 0.03$ for instance.
   We find that $\delta r$ reads $[2.4, 1.4, 0.7]\times 10^{-3}~R$ for $kR = [2, 0.7, 0.3]$.
   These values of $\delta r$ are consistently far below $l$, let alone $r_{\rm i}$.
   However, the decrease of $|\mathd \ct^2/\mathd r|_{r=\rc}$ with decreasing $kR$ is
       substantial.
   }.
When it comes to the $k$-dependence of $X$, the $k$-dependence 
   of $(\mathd \ct^2/\mathd r)_{r = \rc}$ and hence that of $\Lambda$
   somehow counteract the $k^2$ term.
While $X$ remains largely determined by $k^2$, this subtlety 
   means that the critical $kR \approx 0.7$ is specific to
   the equilibrium configuration adopted in this study. 
The insensitivity of the critical $kR$ to $l/R$, on the other hand, 
   simply follows from
   the $l/R$-insensitivity of $\omgR/k$ for a given $k$
   (see Figure~\ref{fig_disp_lR}b). 

When the TB approximation is not that satisfactory ($l/R > (l/R)^{\rm TB})$,
   one needs to employ the resistive computations
   to examine the relative importance of the resonances.
While isolating one resonance is no longer possible, 
   such an examination remains possible 
   because the importance of the cusp relative to the \Alf\ resonance
   is quantified by $\hatFc/\hatFA$ in view
   of Equation~\eqref{eq_linMHD_ener_cons_Eig_shell}.
At this point, we need to clarify how $X$ in Equation~\eqref{eq_X_TB}
   relates to $\hatFc/\hatFA$ to avoid possible confusion. 
For this purpose, we recall that the governing equations on the primitive variables
   were sufficient for \citet{1991SoPh..133..227S} to develop the mathematical prototype 
   that enables the derivation of the TB dispersion relation \eqref{eq_TB} where
   $X$ is involved. 
We recall further that Equation~\eqref{eq_linMHD_ener_cons_Eig_shell}
   is valid regardless of the TB approximation.   
It is just that $\hatFc/\hatFA$ can be explicitly expressed by 
   Equation~\eqref{eq_X_TB} when the TB limit holds.
To see this, we use Equations~\eqref{eq_def_Fc} and \eqref{eq_def_FA}
   to arrive at
\begin{eqnarray}
\displaystyle 
  \frac{\hatFc}{\hatFA}
 = \frac{\left\{
            r {\rm Re}(\ptottilde^* \tilde{v}_r)\right\}_{\rm c}}
	    {\left\{
            r {\rm Re}(\ptottilde^* \tilde{v}_r)\right\}_{\rm A}}~, 
            \label{eq_X_byF}           
\end{eqnarray}   
    where the braces are recalled to evaluate the difference
    of some quantity across a dissipation layer (DL) of finite width. 
In the TB limit, the DLs are thin as well and $\ptottilde$ is continuous
    across a DL \citep[see][for details]{2011SSRv..158..289G}.    
We can then assume that
    $\rcminus \approx \rcplus \approx \rc$, $\rAminus \approx \rAplus \approx \rA$,
    and replace the braces
    with a pair of square parentheses. 
Further assuming that $r\ptottilde$ is the same at the two resonances, 
    one finds that Equation~\eqref{eq_X_byF} evaluates to    
    $[\tilde{v}_r]_{r = \rc}/[\tilde{v}_r]_{r = \rA}$ or equivalently 
    $[\tilde{\xi}_r]_{r = \rc}/[\tilde{\xi}_r]_{r = \rA}$
    because $\tilde{v}_r = -i \omega \tilde{\xi}$.
The general expression \eqref{eq_X_byF} therefore 
    falls back to Equation~\eqref{eq_X_TB}.   
In other words, the relative importance $\hatFc/\hatFA$  
   is fully determined by
   the jumps in the radial speed in the TB limit.
When the TB limit is not that good an approximation,    
   the relative importance can be roughly measured
   by how $\{\tilde{v}_r\}_{\rm c}$ compares with $\{\tilde{v}_r\}_{\rm A}$
   if $\{r \ptottilde\}_{\rm c}$ is not that different
   from $\{r \ptottilde\}_{\rm A}$.
     
Figure~\ref{fig_eigfunc} presents the radial profiles 
   of the Fourier amplitudes of some relevant perturbations
   for an equilibrium profile with 
   a TL-width-to-radius ratio ($l/R$) of $0.2$.
The vertical dash-dotted lines mark the boundaries 
   of the TL.    
The top row represents the Eulerian perturbation to 
   total pressure ($\ptottilde$), while the rest pertain to 
   the radial    ($\tilde{v}_r$), 
       azimuthal ($\tilde{v}_\theta$),
   and axial ($\tilde{v}_z$) speeds. 
The eigen-functions are normalized such that $\ptottilde = 1$ 
   where $|\ptottilde|$ attains its maximum.
For any eigen-function, 
   we use the black and red
   curves to plot its real and imaginary parts, respectively.    
Two axial wavenumbers ($kR$) are examined, one being $0.3$ (the left column)
   and the other being $2$ (right).
Note that the resistive solutions for different values of $kR$ 
   pertain to different magnetic Reynolds numbers ($\Rm$).
To be specific, 
   for each $kR$, we choose an $\Rm$ that is barely inside
      the resonant regime.  
The reason for doing this is that when $\Rm$ is too large,
   $\tilde{v}_r$ in the DLs may be too oscillatory 
   for us to estimate the relative importance of the resonances.    
Actually the oscillatory behavior can be seen in the lower two
   rows in Figure~\ref{fig_eigfunc}, from which the cusp (\Alf) resonance
   can be readily told by the strong oscillations in $\tilde{v}_z$
   ($\tilde{v}_\theta$).
The so-called $1/s$-singularities are also clearly visible.   
The presence of both resonances is further seen in the plots
   for $\tilde{v}_r$,
   where both the jumps across resonances
     (e.g., the black curves)
   and logarithmic singularities (e.g., the red curves)
     are evident
   \footnote{These singular profiles are
   only apparent because the resistivity is nonetheless finite
   \citep[see the review by][for details]{2011SSRv..158..289G}.}. 
In contrast, the curves for $\ptottilde$ are essentially continuous,
   as expected in the case of weak damping.         
Comparing the two columns, one sees a change in the behavior of 
   the magnitude of $\{\tilde{v}_r\}_{\rm c}$
   relative to $\{\tilde{v}_r\}_{\rm A}$.
While $\{\tilde{v}_r\}_{\rm c}$ is less strong than $\{\tilde{v}_r\}_{\rm A}$
   for $kR = 0.3$, the opposite happens for $kR = 2$. 
One then expects that the cusp resonance is less important
   in the former case, but is stronger in the latter.  
   
Equation~\eqref{eq_linMHD_ener_cons_Eig_shell} can be used to perform
   a more quantitative evaluation
   of the relative importance of the resonances 
   as a function of the axial wavenumber $kR$. 
For this purpose, we adopt a fixed $l/R = 0.2$ and
   plot the ratios of the relevant terms 
   in Figure~\ref{fig_reson_damp_k_energetics}
   in a format nearly identical to Figure~\ref{fig_reson_damp_l_energetics}.
One sees that the sum of the ratios, 
   $(\hatFc+\hatFA+\hat{S}_{\rm ideal})/\hat{E}$, agrees with the dashed curve
   representing the values of $-2\omgI$, meaning that
   the computed eigen-functions are remarkably accurate.
More importantly, one sees that the cusp resonance surpasses the \Alf\ one
   at $kR \sim 1$ in terms of the importance for damping the SSKMs.         
That the cusp resonance dominates for large $kR$ is nothing 
   unexpected. 
What Figures~\ref{fig_disp_lR} and 
   \ref{fig_reson_damp_k_energetics} suggest
   is that the cusp resonance may not be
   the dominant resonance throughout the entire $kR$ range
   that is likely to be observationally relevant. 
    
The relative importance of the two resonances aside, 
   one may question how efficient 
   the resonant damping of SSKMs can be.
For this purpose, we recall that       
   for each $kR$, 
   there exists an $l/R$ that optimizes the resonant damping
   (see Figure~\ref{fig_disp_lR}a; 
   see also \citeauthor{2019PhPl...26g0705Y}~\citeyear{2019PhPl...26g0705Y}
   for a relevant study on
   coronal kink waves).
Let $\lRmin$ denote this $l/R$, and $\tauPmin$ denote the value
   that the damping-time-to-period ratio attains therein. 
Figure~\ref{fig_tauPmin} presents both $\lRmin$ (the red curve)
   and $\tauPmin$ (green) as a function of $kR$.     
One sees that $\lRmin$ decreases with $kR$, reaching values below 
   $0.02$ when $kR \gtrsim 16$. 
This inspired us to examine the TB expectations at a given $\lRmin$,
   and the corresponding results are shown by the blue curve 
   labeled $(\tau/P)^{\rm TB}$.
Despite that $\lRmin$ eventually tends to be small,
   the TB expectations are seen to consistently overestimate
   the resonant damping rather considerably. 
In other words, at a given $kR$, 
   the TB limit holds only for $l/R$ that is even smaller than $\lRmin$
   (see Figure~\ref{fig_disp_lR}a).       
One further sees that when $kR$ increases, 
   the resistive value for $\tauPmin$ 
   rapidly decreases at first but levels off
   when $kR$ exceeds, say, $10$. 
The asymptotic value that $\tauPmin$
   attains at $kR = 30$ is merely $\sim 70$.
From this we conclude that, in the observationally relevant range of $kR$, 
   resonant absorption does not play an important role for damping
   the SSKMs in our equilibrium setup.

\section{SUMMARY}
\label{sec_conc}
Motivated by the considerable interest in 
    solar photospheric collective modes,
    we have examined the damping of slow surface kink modes (SSKMs)
    in photospheric waveguides with equilibrium quantities 
    representative of pores and sunspot umbrae. 
We adopted a cylindrical equilibrium configuration comprising
    a uniform interior, a uniform exterior, and
    a transition layer (TL) that continuously connects the two. 
By formulating the relevant eigenvalue problem
    in the framework of resistive, linear MHD, we were allowed
    to simultaneously account for two mechanisms for damping SSKMs
    in a self-consistent manner,
    one being the Ohmic resistivity, and the other being
    the resonant absorption of SSKMs in the cusp and \Alf\ continua. 
We find that the diversity of the geometric and physical parameters of
    photospheric waveguides, wrapped up in the magnetic Reynolds number ($\Rm$),
    means that the relative importance of the two mechanisms     
    needs to be assessed on a case-by-case basis.
At realistically small values of $\Rm$, the Ohmic resistivity
    plays a far more important role, resulting
    in damping-time-to-period-ratios $\tau/P \sim 10$
    in the parameter range we examine.
In contrast, resonant absorption is the only damping mechanism
    for some realistically large $\Rm$.
In this latter case, the cusp resonance in general dominates
    the \Alf\ one except for very long axial wavelengths. 
Regardless, in the observationally relevant wavelength range,
    the resonant damping rate turns out to be negligible, leading to
    values for $\tau/P$ to
    consistently exceed $\sim 70$.

When SSKMs are damped solely by resonant absorption, we have 
    compared the damping rates obtained by the resistive approach
    with those found by solving the explicit dispersion relation valid
    in the thin-boundary (TB) limit.
In general we find that TB values agree with the resistive ones 
    only when the transition-layer-width-to-radius ratios ($l/R$)
    are much smaller than the nominal values for the TB limit to apply.
This is particularly true for short axial wavelengths, which are
    more observationally relevant.
From the practical point, this means that when the resonant damping of SSKMs
    is of interest, one needs to adopt approaches such as the resistive one, despite 
    that the TB limit proves much less computationally expensive.

Before closing, let us name a limited number of intricacies that are likely
    inherent 
    in theoretical studies on photospheric modes by contrasting
    our findings with representative coronal studies on radial fundamental kink modes.
One, the range of $\Rm$ where resonant absorption is expected
    to dominate for damping collective modes.  
Drawing experience from the coronal studies by, say, \citet{2006ApJ...642..533T} 
     and \citet{2016SoPh..291..877G}, one is inclined to propose that
     photospheric SSKMs are damped primarily by resonant absorption
	 given that $\Rm$ tends to exceed $\sim 10^4$. 
However, this is not the case.
Two, the range of applicability of the TB limit regarding the resonant damping of 
    collective modes.
While the TB limit is constructed
     to hold only when $l/R$ is small,
     the range of its applicability is drastically different in photospheric cases
     from that in coronal cases.
In coronal studies, the TB expectations
     for the resonant damping rates of 
     kink modes tend to be a good approximation well
     beyond the nominal range of applicability      
     \citep[e.g.,][]{2004ApJ...606.1223V,2014ApJ...781..111S}.
In particular, Figure~1 in the latter study indicates that
     the true values for $\tau/P$ deviate from the TB ones
     by no larger than $80\%$ even when $l/R$ reaches two, the maximum
     allowed by the equilibrium configuration.     
And this deviation is $\lesssim 20\%$ when $l/R \lesssim 0.2$.
This is not the case for photospheric SSKMs, for which 
     the relative deviation
     can readily exceed unity even for $l/R$ as small as, say, $0.1$.
We take these two intricacies as strengthening rather than weakening the need
     to further the theoretical examinations on photospheric modes. 
For instance, the second intricacy means that other approaches like the one 
     outlined for coronal kink modes by \citet{2013ApJ...777..158S} 
     can readily find photospheric applications.
Likewise, the first intricacy means that the present
     study on the non-ideal mechanisms for damping SSKMs
     needs to be taken further.
Drawing experience from the study by \citet{2020ApJ...897..120G}
     on the damping of photospheric slow surface sausage modes (SSSMs)
     , we note that the Ohmic resistivity can be so efficient 
     that SSSMs are damped within a fraction of the wave period 
     even if $\Rm$ remains $\gtrsim 10^4$ (Figure~3 therein).
While this was found for photospheric waveguides with 
     piece-wise constant transverse distributions of the equilibrium quantities, 
     there seems to be no reason to believe that
     photospheric waveguides cannot be transversely
     structured this way.  

Some remarks on the caveats of our approach are also necessary, 
    for which purpose we start by noting that in reality the solar
    photosphere is structured
    in a way much more complicated than modeled here. 
In fact, this complication is to such an extent that we have to restrict
    the following discussions
    to only one additional factor that is expected to be important
    in the wave dynamics. 
This factor, the effect of gravity, was neglected in our study
    from the outset.
As a consequence, essentially only three spatial scales were
    relevant, namely the waveguide radius ($R$),
    the TL width ($l$), and the axial wavelength ($\lambda = 2\pi/k$).
Introducing gravity considerably complicates the problem, some aspects of which
    were summarized by \citet[][Chapters~9 to 11 and references therein]{roberts_2019CUP}.
For our purpose, it suffices to assume that the gravitational acceleration
    is aligned with the waveguide axis. 
A number of additional spatial scales
    become relevant
    even if we assume that
    neither the equilibrium temperature nor the mean molecular weight 
    varies much over photospheric heights.
These scales characterize how rapid the equilibrium pressure and magnetic field strength
    vary.
Let $H$ denote the shortest one among these scales.    
Given the axial stratification, 
    the axial wavelength ($\lambda$) cannot be unambiguously defined, but 
    nonetheless can be understood as the spatial scale characterizing the axial variation
    of the perturbations. 
For our results to apply, the condition $\lambda \ll H$ should be satisfied
    but proves quite demanding for the majority of our computations in view of the
    small values of $H$ at photospheric temperatures. 
It then follows that, in principle our results apply only to 
    ER83-like photospheric waveguides, by which we mean the idealized configurations
    where the equilibrium quantities are structured only transversely
    and the involved characteristic speeds follow some canonical ordering. 
Note that here we distinguish between the ER83-like and ER83 equilibria 
    by one key difference, namely the transverse structuring is allowed to be continuous in the former as realized by a non-vanishing $l$ in this study.
With this distinction we note that all the referenced
    observational studies on photospheric modes have adopted the ER83 framework
    to aid the mode identification and/or 
    the evaluation of energetics. 
Focusing on wave damping, we believe that our computations contributed to the
    better understanding
    of photospheric collective modes, ``photospheric modes'' in the sense
    as practiced in available observational and theoretical efforts.
In addition, our results can be of help for
    further theoretical/numerical studies that address additional realistic effects
    on photospheric modes.  
For instance, if one adopts a theoretical framework that extends ours by accounting for
    gravity, then our results can be used to validate the pertinent computations by
    specializing to the cases where $\lambda \ll H$.
We believe that such a validation is warranted, given that such further
    theoretical developments are necessarily more involved than 
    presented here.
That said, we concede that it is difficult to assess how well
    our results carry over to the extremely complicated photospheric structures
    \citep[see also][the introduction sections in Chapters 9 and 10]{roberts_2019CUP}.

With the above-mentioned caveats in mind, let us discuss whether our results
    can help constrain the likely damping mechanism(s) for an identified SSKM.
We start by noting that, in the case of weak damping, 
    the damping-time-to-period ratios $\tau/P$ for temporally damped modes
    are nearly identical to the damping-length-to-wavelength ratios $L_{\rm D}/\lambda$
    for spatially damped ones if the wave dispersion is weak
    \citep[see e.g.,][Equation~40]{2010A&A...524A..23T}.
In view of the extremely weak dispersion of SSKMs, this means that the
    following discussions apply to SSKMs that experience either temporal
    or spatial damping. 
For the ease of discussion, let us slightly generalize
    Equation~\eqref{eq_omega_formal} by counting all possible physical quantities
    that may influence the dispersive properties of SSKMs in our framework. 
Some straightforward dimensional analysis yields that the complex-valued frequency
    $\omega$ can be expressed as
\begin{eqnarray}
\label{eq_omega_formal_generalized}
\displaystyle 
  \frac{\omega R}{v_{\rm Ai}} 
= {\cal H}
   \left(
      \frac{\csi}{\vai}, \frac{\cse}{\vai}, \frac{\vae}{\vai}, \frac{l}{R}, R_{\rm m};
      kR
   \right)~,
\end{eqnarray}    
    where the function ${\cal H}$ is dictated by the eigen-value problem. 
Note that the arguments on the RHS
    of Equation~\eqref{eq_omega_formal_generalized} 
    are all dimensionless, and are independent from one another.
One may further formally express $\tau/P$ (or equivalently $L_{\rm D}/\lambda$) as
\begin{eqnarray}
\label{eq_tauP_formal_generalized}
\displaystyle 
  \frac{\tau}{P} 
= {\cal L}
   \left(
      \frac{\csi}{\vai}, \frac{\cse}{\vai}, \frac{\vae}{\vai}, \frac{l}{R}, R_{\rm m};
      kR
   \right)~.
\end{eqnarray}
To proceed, let us assume that the set of 
   $[\csi/\vai, \cse/\vai, \vae/\vai]$ is measurable.
Our Figure~\ref{fig_tauPmin} then indicates that 
    $\tau/P$ cannot be lower than 
    the asymptotic value $(\tau/P)_{\rm min, asymp}$ that the green curve
    attains at large $kR$, provided that
    resonant absorption is the sole damping mechanism.
Conversely, the wave damping cannot be attributed to resonant absorption
    if $\tau/P$ is observed to be smaller than $(\tau/P)_{\rm min, asymp}$, 
    which depends only on $[\csi/\vai, \cse/\vai, \vae/\vai]$.
In this case, one is allowed to rule out resonant absorption
    even without knowing any quantity in the set $[l/R, \Rm, kR]$.
The situation is much more involved when the observed $\tau/P$
    exceeds $(\tau/P)_{\rm min, asymp}$, and one needs to survey
    the combinations of $[l/R, \Rm, kR]$ (see Figure~\ref{fig_disp_Rm}).
However, it remains possible to say a few words on the likely
    damping mechanism(s).
For this purpose, we further assume that 
    $l/R$ and $kR$ are known.
Note that it is non-trivial to evaluate the electric resistivity ($\eta$) 
    in a photospheric medium
    \citep[e.g.,][]{1993ASPC...46..465W,2012ApJ...747...87K,2016ApJ...832..195N}.
Nonetheless, one can rather safely assume that the resulting $\Rm$ lies in
    the range pertaining to the shaded area in Figure~\ref{fig_disp_Rm}, given that
    this range encompasses the rather extreme values of the waveguide radius ($R$).
With a known set of $[\csi/\vai, \cse/\vai, \vae/\vai, l/R; kR]$, 
    one may construct the $\Rm$-dependence of $\tau/P$, thereby finding
    the range of $\tau/P$ allowed by the range of $\Rm$.
Now two situations arise.
If the observed $\tau/P$ lies in this range, then one is allowed
    to constrain the possible $\Rm$.
If the measured $\tau/P$ is outside this range, then
    the wave damping can be attributed to neither the Ohmic resistivity
    nor the resonant absorption.
To name but one additional mechanism, we note that we have adopted the values for
    the electric conductivity from \citet{1983SoPh...84...45K}.
As such, only the Ohmic resistivity due to electron-neutral and electron-ion collisions
    was addressed.
In reality, the Cowling resistivity due to ion-neutral collisions
    may be substantially stronger \citep[e.g.,][Figure~1]{1993ASPC...46..465W}.
A detailed study on the effects of the Cowling resistivity on photospheric modes
    is certainly warranted, but is nonetheless left for a future work.

\acknowledgments
We thank both referees for their comments, which helped improve
    the manuscript substantially.
This research was supported by the 
    National Natural Science Foundation of China
    (BL: 41674172, 11761141002, 41974200; SXC:41604145, HY:41704165).
TVD was supported by the European Research Council (ERC) under 
   the European Union's Horizon 2020 research 
   and innovation programme (grant agreement No 724326) 
   and the C1 grant TRACEspace of Internal Funds KU Leuven. 
This research benefited greatly from the discussions at ISSI-BJ.
This research was also supported by Shandong University
   via grant No 2017JQ07.


\bibliographystyle{aasjournal}
\bibliography{seis_generic}



\clearpage
\begin{figure}
\centering
 \includegraphics[width=.6\columnwidth]{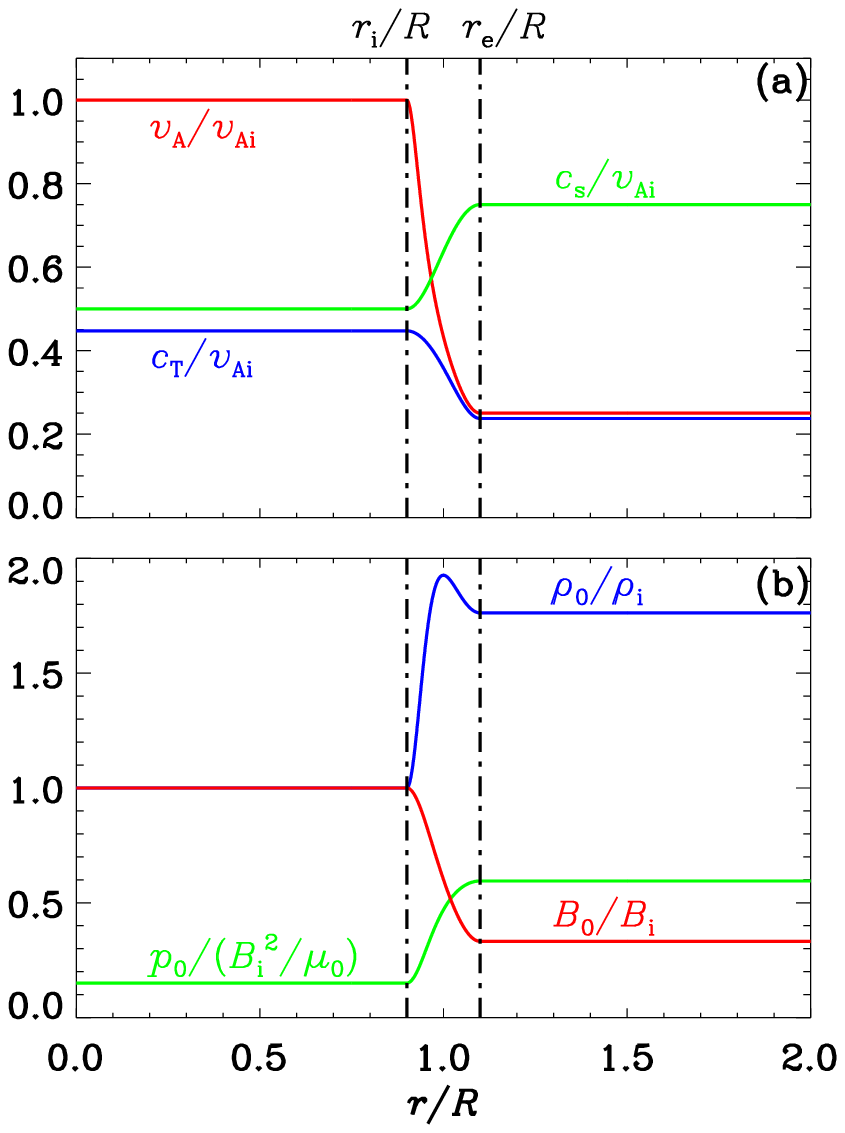}
 \caption{Transverse profiles for 
 (a) the characteristic speeds and 
 (b) primitive variables of a photospheric waveguide
     representative of pores and sunspot umbrae.
 These profiles are generated by specifying the transverse distributions of
     the adiabatic sound ($c_{\rm s}$) and cusp ($c_{\rm T}$) speeds, for which
     a continuous transition layer (TL) connects a uniform interior to a uniform exterior
     (see Equation~\ref{eq_profile_cs_ct}).
 This TL is located between $r_{\rm i} = R-l/2$ and $r_{\rm e} = R+l/2$, where
     $R$ is the mean waveguide radius and $l$ the TL width.
 For illustration purposes, here      
     $l/R$ is arbitrarily chosen to be $0.2$.      
 }
 \label{fig_EQprofile}
\end{figure}

\clearpage
\begin{figure}
\centering
 \includegraphics[width=0.6\columnwidth]{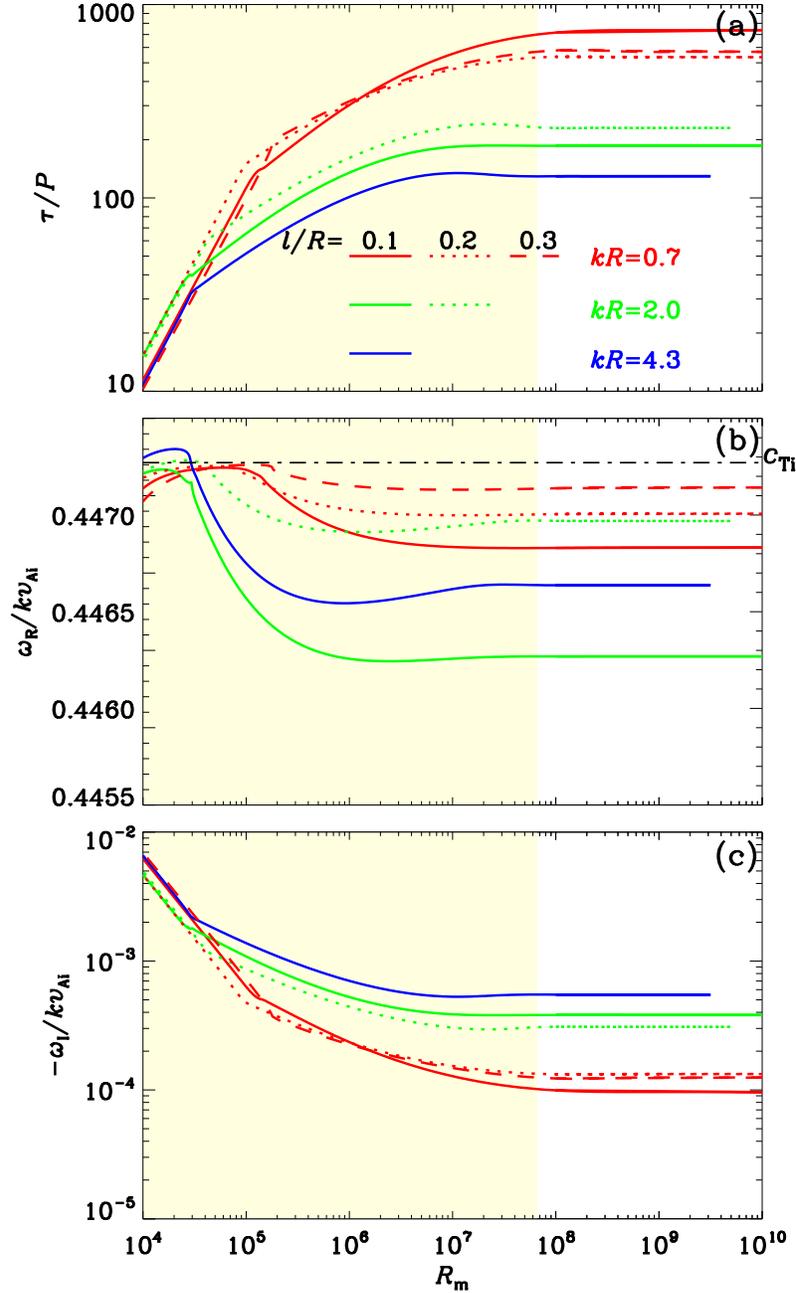}
 \caption{
 Dependence on the magnetic Reynolds number ($R_{\rm m}$) 
    of the dispersive properties
    of slow surface kink modes in a photospheric
    waveguide.
Plotted are (a) the damping-time-to-period ratio $\tau/P$,
        (b) the axial phase speed $\omgR/k$,
    and (c) the damping rate $\omgI$ in units of $k\vai$.
A number of combinations of
    the transition layer width ($l/R$)
    and axial wavenumber ($kR$) are examined as labeled.
 The area shaded yellow corresponds to the range of $R_{\rm m}$ 
    that is likely to be relevant for pores and sunspot umbrae.
 See the text for details.
}
 \label{fig_disp_Rm}
\end{figure}

\clearpage
\begin{figure}
\centering
 \includegraphics[width=0.8\columnwidth]{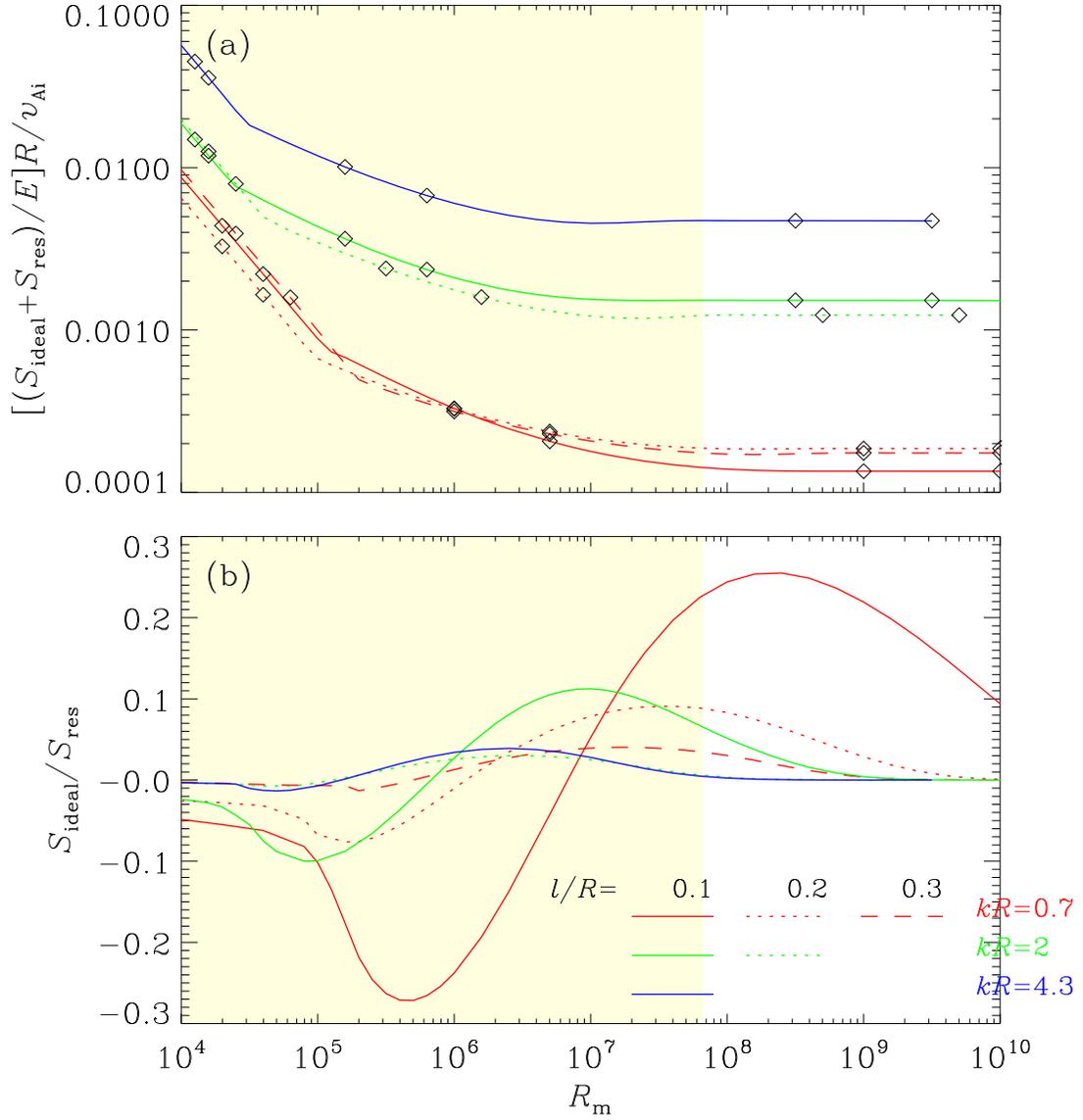}
 \caption{
 Dependence on the magnetic Reynolds number ($R_{\rm m}$) 
    of the terms involved in the energy balance
    (see Equation~\ref{eq_linMHD_ener_cons_Eig_EntireVol}
    and the associated definitions).
 A number of combinations of
    the transition layer width ($l/R$)
    and axial wavenumber ($kR$) are examined as labeled.
 The area shaded yellow corresponds to the range of $R_{\rm m}$ 
    that is likely to be relevant for pores and sunspot umbrae.
 The diamonds represent $-2\omgI$, taken from 
    a small subset of the values presented in Figure~\ref{fig_disp_Rm}c.     
 See the text for details.
}
 \label{fig_enerbal_Rm}
\end{figure}

\clearpage
\begin{figure}
\centering
 \includegraphics[width=0.6\columnwidth]{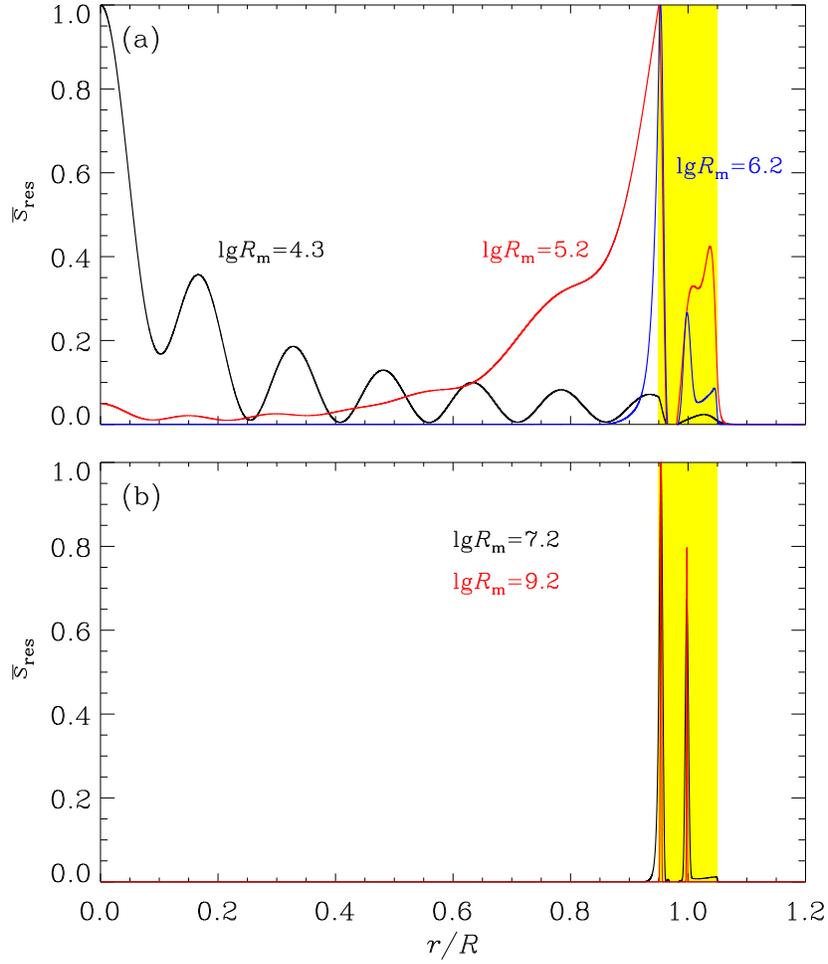}
 \caption{
 Radial distributions of the $\bar{s}_{\rm res}$ term for a number of
     representative values of the magnetic Reynolds number ($R_{\rm m}$)
     at a given pair of $[l/R, kR] = [0.1, 0.7]$. 
 Here $\bar{s}_{\rm res}$ is evaluated by applying the bar operation
     (Equation~\ref{eq_def_bar})
     to the definition of $\sres$ (Equation~\ref{eq_linMHD_sres_final}).        
 The shaded area represents where the equilibrium quantities are nonuniform.     
 See the text for details.
}
 \label{fig_Sres_r_distri}
\end{figure}

\clearpage
\begin{figure}
\centering
 \includegraphics[width=.6\columnwidth]{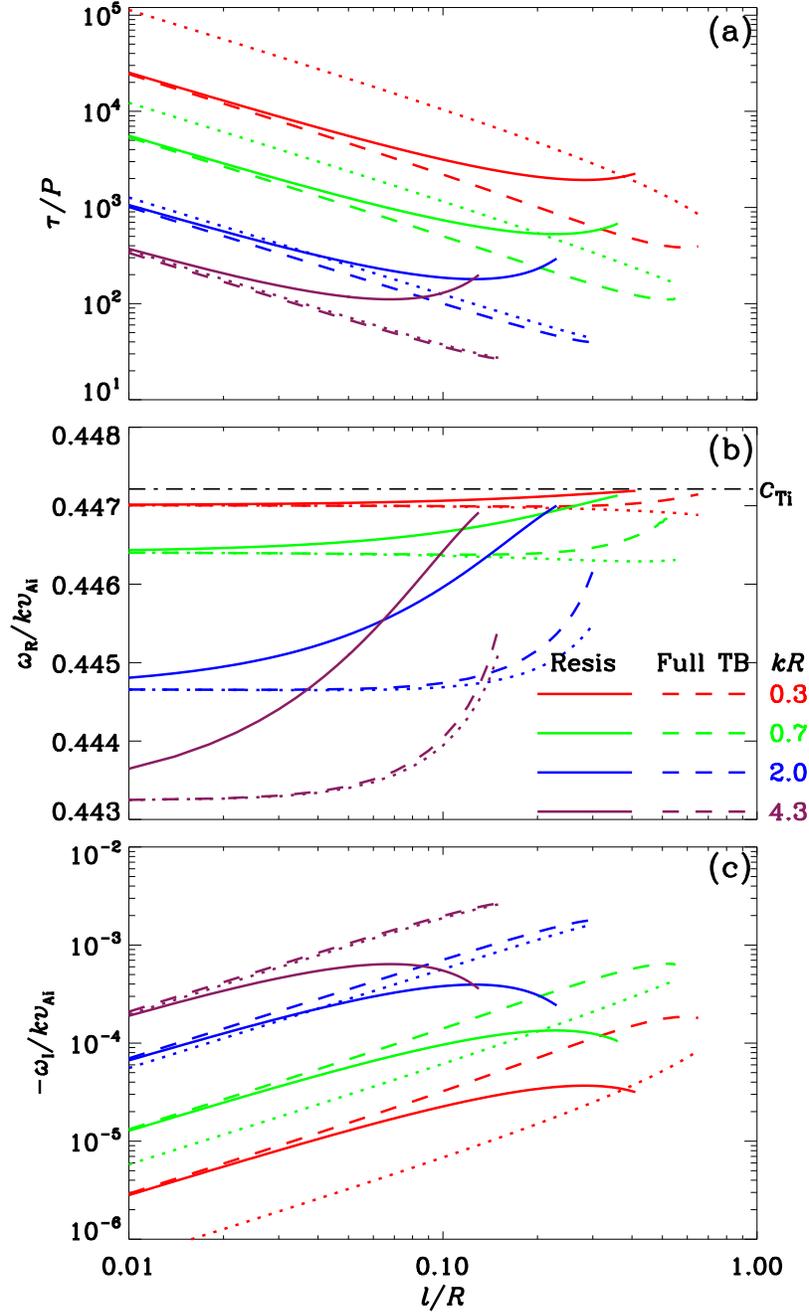}
 \caption{
 Dispersive properties of resonantly damped slow surface kink modes, 
    shown by the dependence on the dimensionless
    layer width ($l/R$) 
    of (a) the damping-time-to-period ratios ($\tau/P$),
       (b) the axial phase speeds ($\omgR/k$),
   and (c) the damping rates ($\omgI$) in units of $k\vai$.
A number of axial wavenumbers ($kR$) are examined as 
   represented by the different colors.
The results from self-consistent resistive computations
   are given by the solid curves. 
For each pair of $[l/R, kR]$, the explicit dispersion relation (Equation~\ref{eq_TB})
   in the thin-boundary limit is also solved.
The dashed curves represent the computations where both the cusp
   and \Alf\ resonances are accounted for, whereas the dotted curves
   take into account only the cusp resonance. 
}
 \label{fig_disp_lR}
\end{figure}

\clearpage
\begin{figure}
\centering
 \includegraphics[width=.95\columnwidth]{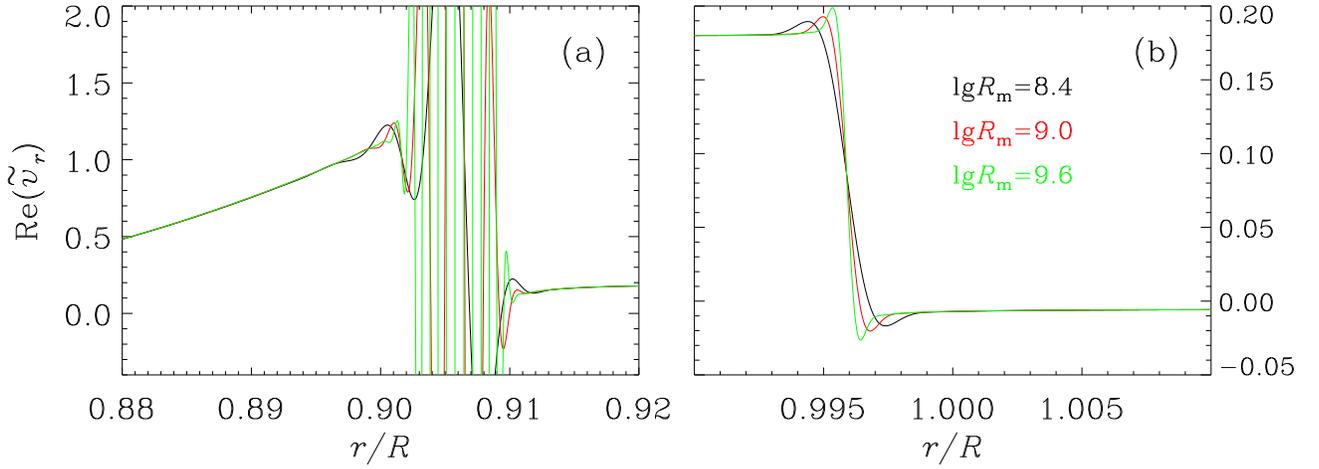}
 \caption{
Radial distributions of the real part of the Fourier amplitude
   of the radial flow speed (${\rm Re}\tilde{v}_r$) 
   in the intervals embracing 
       (a) the cusp,
   and (b) the \Alf\ dissipative layers. 
A number of magnetic Reynolds numbers ($\Rm$)
   are examined as labeled, whereas $[l/R, kR]$ is fixed at $[0.2, 2]$. 
The eigenfunctions are rescaled such that 
   $\ptottilde = 1$ at $r = 1.25~R$.
   } 
\label{fig_reson_Revr_r}
\end{figure}

\clearpage
\begin{figure}
\centering
 \includegraphics[width=.95\columnwidth]{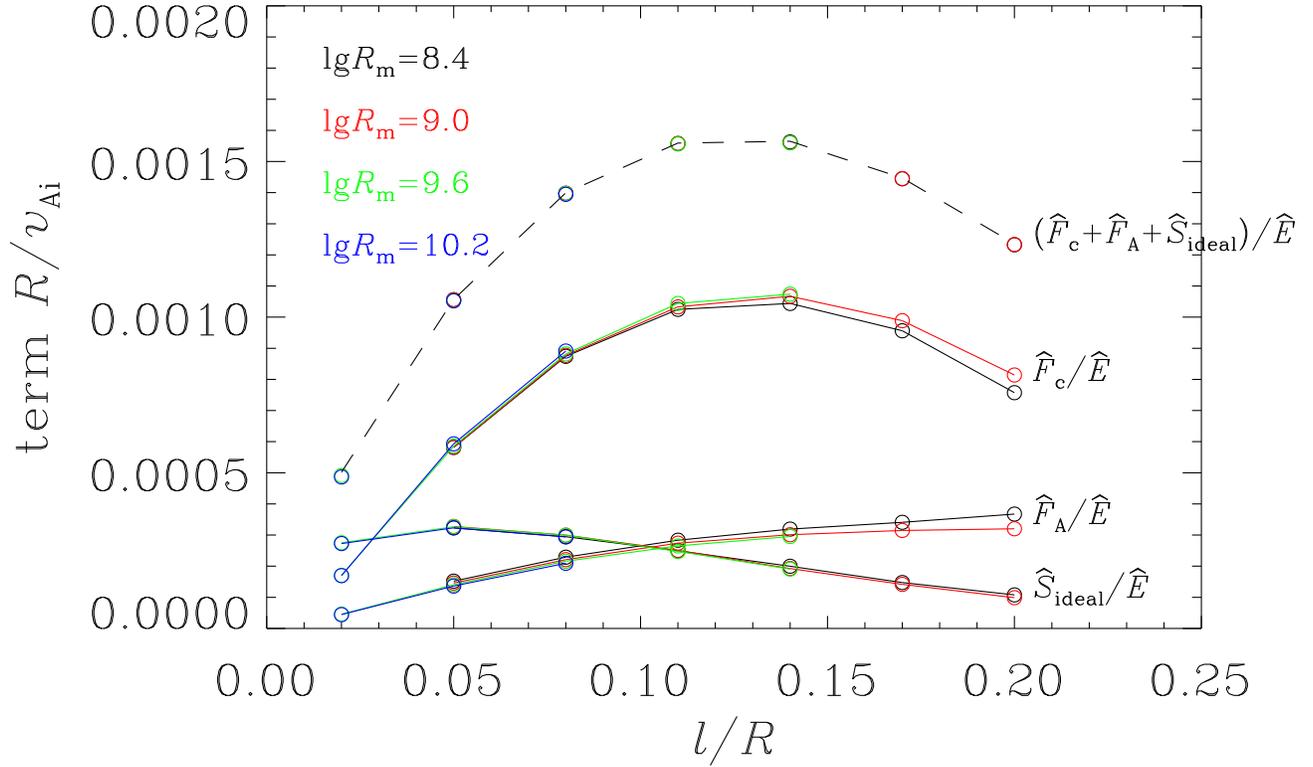}
 \caption{
Dependence on the transition layer width ($l/R$) of 
   the ratios of the hatted terms
   in Equation~\eqref{eq_linMHD_ener_cons_Eig_shell}.
The axial wavenumber is fixed at $kR = 2$.
For a given $l/R$, a number of magnetic Reynolds numbers ($\Rm$)
   are examined to evaluate the hatted terms
   as represented by the symbols of different colors. 
With solid curves we 
   connect the symbols representing $\hatFc$, $\hatFA$, and $\hat{S}_{\rm ideal}$
   for the same $\Rm$ but different $l/R$.
The dashed curve represents $-2\omgI$, the values of which
   are direct outputs of our code.
See the text for details.   
   } 
\label{fig_reson_damp_l_energetics}
\end{figure}

\clearpage
\begin{figure}
\centering
\includegraphics[width=.8\columnwidth]{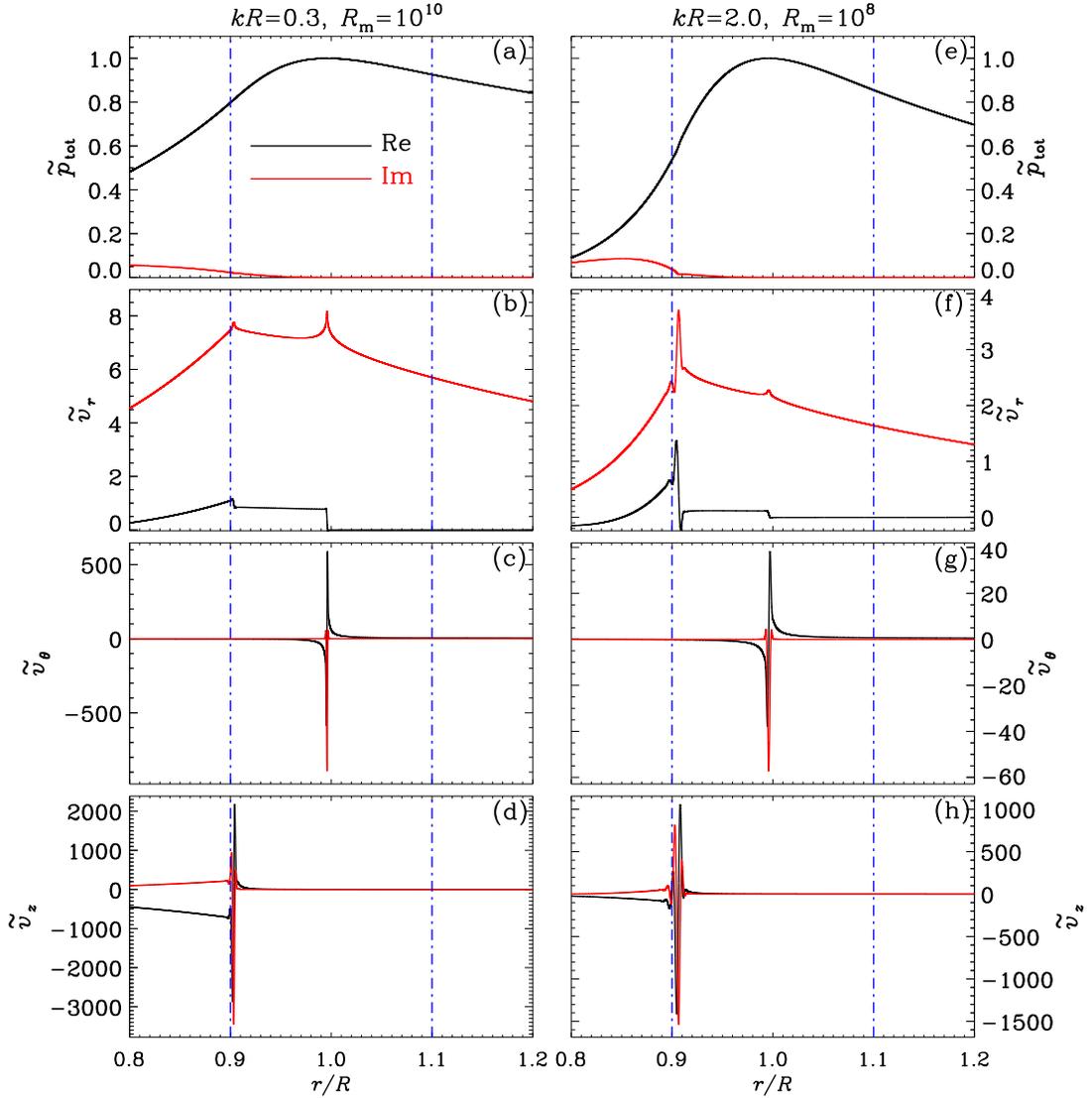}
 \caption{
 Radial profiles of some relevant eigen-functions found in
     self-consistent resistive computations 
     for resonantly damped slow surface kink modes. 
 Two different values for the axial wavenumber are examined, one corresponding
     to $kR = 0.3$ (the left column) and the other to $kR = 2$ (right).
 The vertical dash-dotted lines mark the boundaries of the transition layer,
     which corresponds to a layer-width-to-radius-ratio of $l/R = 0.2$.      
 The black (red) curves correspond to the real (imaginary) parts. 
 The top row pertains to the Eulerian perturbation to total pressure ($\ptottilde$),
     while the rest correspond to
     the radial ($\tilde{v}_r$), azimuthal ($\tilde{v}_\theta$),
     and axial ($\tilde{v}_z$) speeds. 
 The eigen-functions are normalized such that $\ptottilde =1$ where its modulus
     attains maximum.      
 Different values for the magnetic Reynolds number ($\Rm$)
     are adopted in the two columns, see the text
     for details.     
}
 \label{fig_eigfunc}
\end{figure}

\clearpage
\begin{figure}
\centering
 \includegraphics[width=.95\columnwidth]{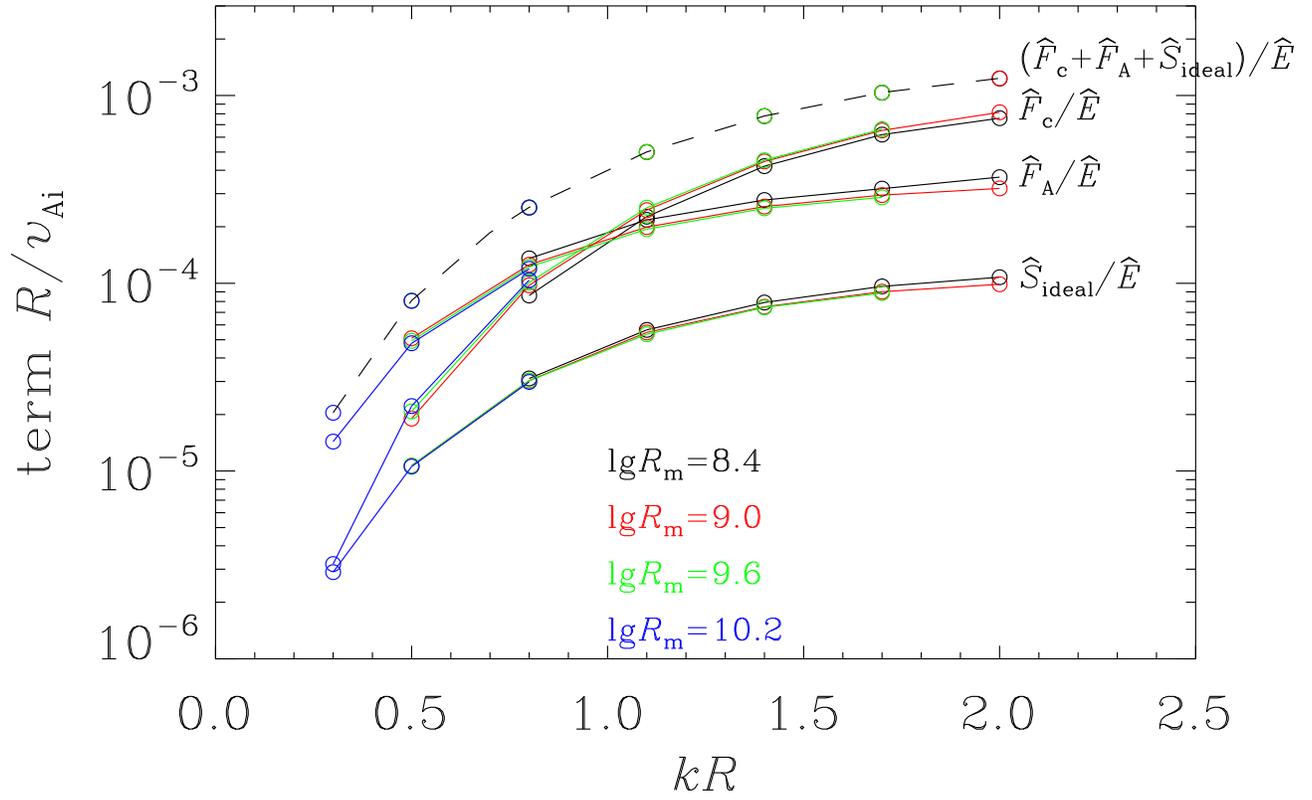}
 \caption{
 Similar to Figure~\ref{fig_reson_damp_l_energetics}
   except that the dependence on the axial wavenumber
   ($kR$) is examined.
 Here the dimensionless transition layer width 
   is fixed at $l/R = 0.2$.   
   } 
\label{fig_reson_damp_k_energetics}
\end{figure}

\clearpage
\begin{figure}
\centering
\includegraphics[width=.8\columnwidth]{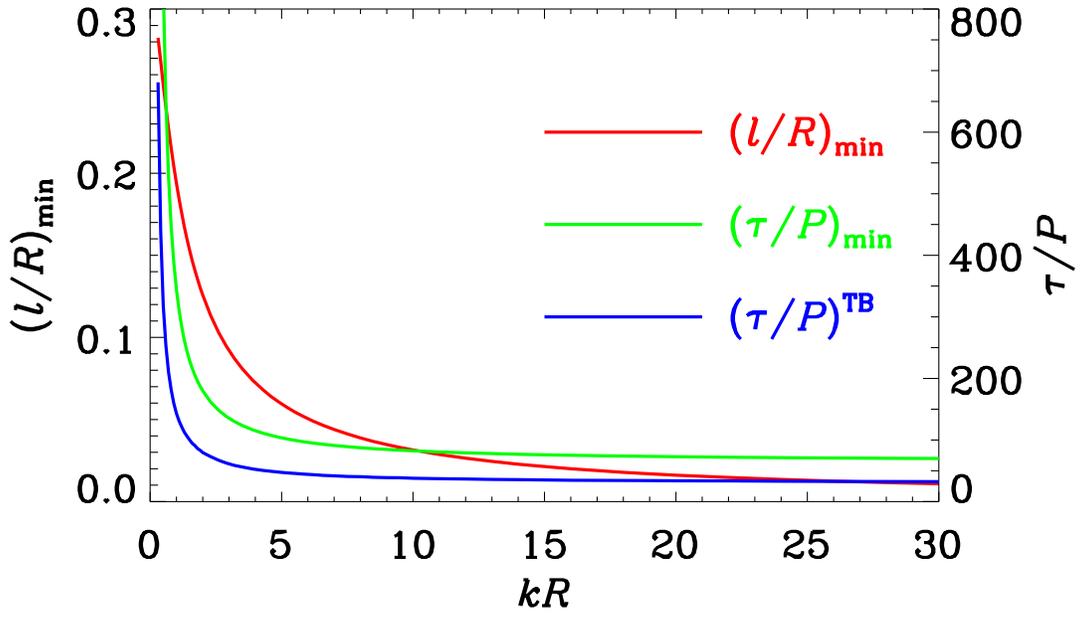}
 \caption{
Dependence on the axial wavenumber ($k$) of
   the optimal damping-time-to-period ratio $(\tau/P)_{\min}$
   and the transition-layer-width-to-radius ratio $(l/R)_{\min}$
   where the optimal damping is reached. 
The values for $(\tau/P)_{\min}$ and $(l/R)_{\min}$
   are derived from self-consistent resistive computations.
At an $(l/R)_{\min}$ thus derived, the explicit dispersion relation
   in the TB limit (Equation~\ref{eq_TB}) is also solved, yielding 
   $(\tau/P)^{\rm TB}$ for comparison.      
See the text for details.
}
 \label{fig_tauPmin}
\end{figure}

\end{document}